\documentclass[aps,prx,twocolumn,superscriptaddress]{revtex4-2}

\usepackage{times}
\usepackage{amsmath}
\usepackage{amssymb}
\usepackage{graphicx}
\usepackage{subfigure}
\usepackage{braket}
\usepackage{bm}
\usepackage{multirow}
\usepackage{hyperref}
\usepackage{color}%

\begin{document}

\title{Einstein-de Haas Effect of Topological Magnons}

\author{Jun Li}
\affiliation{State Key Laboratory of Optoelectronic Materials and Technologies, School of Physics, Sun Yat-Sen University, Guangzhou 510275, China}

\author{Trinanjan Datta}
\email[Corresponding author:]{tdatta@augusta.edu}
\affiliation{Department of Chemistry and Physics, Augusta University, 1120 15$^{th}$ Street, Augusta, Georgia 30912, USA}
\affiliation{State Key Laboratory of Optoelectronic Materials and Technologies, School of Physics, Sun Yat-Sen University, Guangzhou 510275, China}

\author{Dao-Xin Yao}
\email[Corresponding author:]{yaodaox@mail.sysu.edu.cn}
\affiliation{State Key Laboratory of Optoelectronic Materials and Technologies, School of Physics, Sun Yat-Sen University, Guangzhou 510275, China}

\date{\today}

\begin{abstract}
We predict the existence of Einstein-de Haas effect in topological magnon insulators. Temperature variation of angular momentum in the topological state shows a sign change behavior, akin to the low temperature thermal Hall conductance response. This manifests itself as a macroscopic mechanical rotation of the material hosting topological magnons. We show that an experimentally observable Einstein-de Haas effect can be measured in the square-octagon, the kagome, and the honeycomb lattices. Albeit, the effect is the strongest in the square-octagon lattice. We treat both the low and the high temperature phases using spin wave and Schwinger boson theory, respectively. We propose an experimental set up to detect our theoretical predictions. We suggest candidate square-octagon materials where our theory can be tested.
\end{abstract}

\maketitle

\section{Introduction}\label{secI}
The Einstein-de Haas (EdH) effect, predicted in 1908 by the Nobel Laureate Owen W. Richardson, was the test of a very fundamental concept --- did circulating electrons give rise to magnetic moments?~\cite{PhysRevSeriesI.26.248}. Seven years later, Albert Einstein and Wander J. de Haas experimentally demonstrated the possibility of transferring magnetic to mechanical angular momentum~\cite{Einstein1915}. Subsequently, their experimental initiative was perfected by Stewart~\cite{Stewart1918}. The transfer of momentum between electron spins and lattice degrees of freedom in a freely suspended magnetized body is a direct outcome of the angular momentum conservation principle. The rotation can be induced by an externally applied transient magnetic field or temperature. The reverse effect, generation of a magnetic moment by mechanical rotation, called the Barnett effect has also been realized~\cite{BarnettPhysRev.6.239}. The very first gyromagnetic ratio measurements were performed using the EdH effect. At present, $g$-factor measurements performed using the EdH technique provide a more accurate value compared to electron-spin resonance or ferromagnetic resonance\cite{jaafar2009dynamics}. In addition, the gyromagnetic ratio has important consequences in the fields of quantum electrodynamics~\cite{PhysRev.73.416}, Bose-Einstein condensates~\cite{PhysRevLett.96.080405}, molecular magnetism, nano-magneto-mechanics, spintronics, magnonics\cite{PhysRevApplied.9.024029} and ultra-fast magnetism~\cite{Ganzhorn2016,GaranPhysRevB.99.064428}.

The realization of topology, an abstract mathematical concept, in correlated electronic and magnetic systems has revolutionized the field of condensed matter physics and materials science~\cite{hasan2010colloquium}. In a topological state, the bulk and the boundary may have entirely different properties. For example, in a quantum spin Hall state the bulk is insulating, whereas the edge can support helical transport modes. These novel topological states are robust against external perturbation such as an external magnetic field. In the bulk of the quantum Hall state, the topological number that governs the integer Hall conductance (called Chern number) provides a criterion to classify the different topological phases which otherwise cannot be related by a continuous mapping~\cite{hatsugai1993edge}. A topological band theory has been developed to explain bulk-boundary correspondence between the edge and the bulk states~\cite{RevModPhys.88.021004}.

Spin-wave excitations (magnons) can be topologically protected. Recently, it has been demonstrated that topological magnons could be transported along the sample edge in a thermal Hall experiment (chiral edge state)~\cite{onose2010observation}. Different from the electrical Hall effect, in its thermal Hall counterpart, current is produced by a temperature gradient rather than an electric field. Thus, a heat, instead of a charge, current is detected. The presence of Dzyaloshinskii-Moriya (DM) interaction mimics the role of time-reversal breaking magnetic field~\cite{onose2010observation}. Current prototypical examples of topological magnon materials include $\text{Lu}_2\text{V}_2\text{O}_7$~\cite{onose2010observation}, $\text{Cu(1,3-bdc)}$~\cite{hirschberger2015thermal}, and $\text{CrI}_3$~\cite{Chen2018TopologicalSE}. Dirac magnons have been proposed in a honeycomb ferromagnet~\cite{BalatPhysRevX.8.011010}. A chiral topological magnon insulator in three dimensions has also been studied~\cite{KovPhysRevB.97.174413}. The thermal Hall effect has been found in various other magnetic systems such as spin liquids~\cite{watanabe2016emergence,kasahara2018unusual}, multiferroics~\cite{ideue2017giant}, antiferromagnets~\cite{doki2018spin,HottaPhysRevB.99.054422,LossPhysRevB.96.224414,Samajdar2019}, and the pseudogap phase of a cuprate superconductor~\cite{grissonnanche2019giant}. Other neutral quasi-particles such as phonons can also exhibit thermal Hall effect ~\cite{strohm2005phenomenological,sugii2017thermal}. Magnons, topological~\cite{BrataasPhysRevLett.122.107201} or otherwise~\cite{PhysRevB.81.214418}, play an important role in device physics, too.

\begin{figure*}[t]
	\centering
	\includegraphics[width=0.98\linewidth]{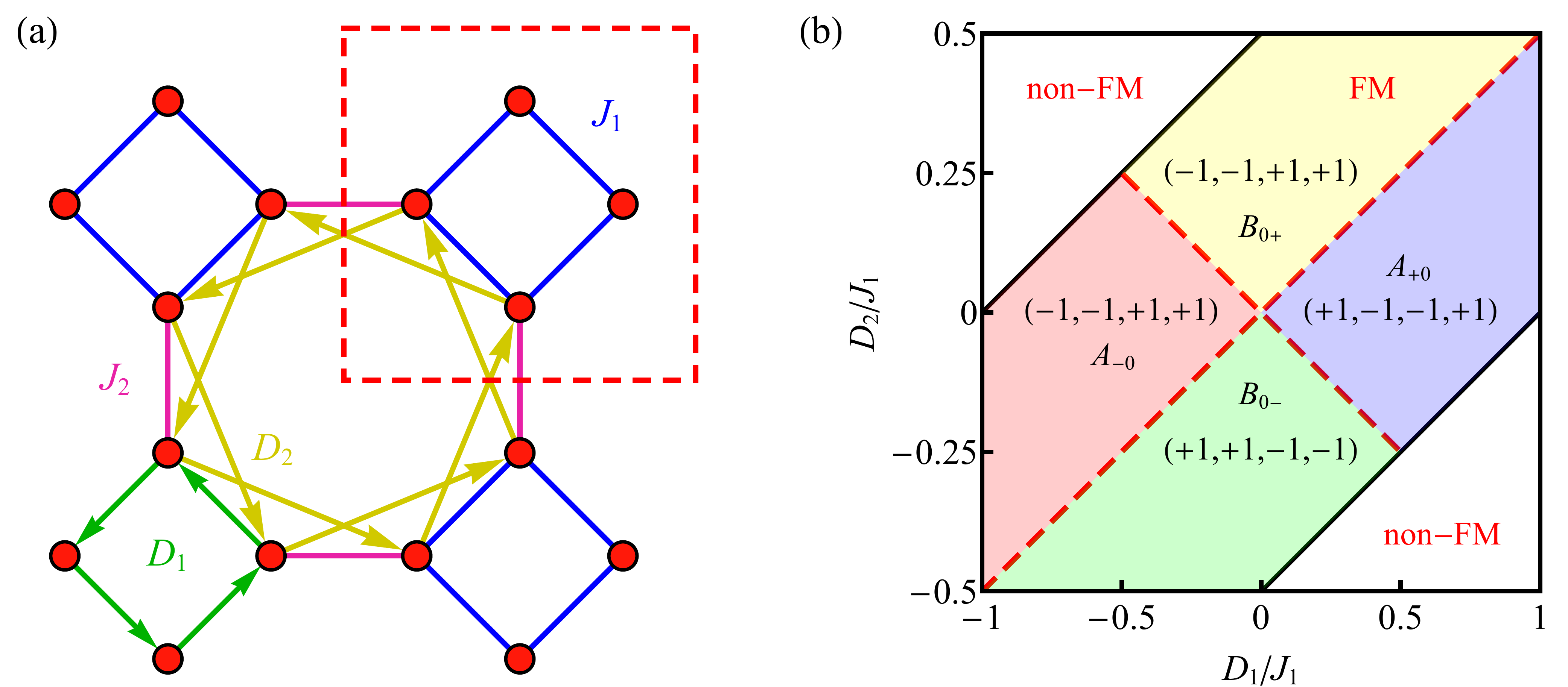}
	\caption{\label{fig:fig1} The square-octagon lattice and the topological phase diagram.~(a)~Lattice and magnetic interactions. $J_1$ and $J_2$ are two types of nearest-neighbor ferromagnetic interactions within cell and between cells, respectively. $D_1$ and $D_2$ denote the nearest and next-nearest neighbor Dzyaloshinskii-Moriya interactions allowed by symmetry, according to the Moriya's rules.~(b)~The topological phase diagram is reported for a generic Chern number set $\{C_1,C_2,C_3,C_4\}$. The index of the Chern number is the same as the energy band order $E_1<E_2<E_3<E_4$. The subscripted $A$ and $B$ symbols classify the various ferromagnetic topological phases. A phase diagram exploring a comprehensive parameter region is reported in the Fig.~\ref{appdfig2}.}
\end{figure*}

In this article, we marry the century old EdH experimental technique with the contemporary concept of topological magnons. We explicitly demonstrate that an EdH effect of topological magnon excitations can be experimentally observed, at least, in three different frustrated magnetic lattices --- square-octagon, kagome, and honeycomb. A temperature evolution map of the angular momentum sign change and the corresponding thermal Hall conductance behavior $ \kappa_{xy} $ for realistic experimental conditions show the validity of our proposal. We provide an estimate of the EdH effect, within a Stewart experiment like set-up, for the square-octagon, the kagome, and the honeycomb lattices. We also compute the EdH response for the honeycomb CrI$_3$ and kagome Cu(1,3 -bdc). We find that the square-octagon geometry has the largest low temperature EdH response within a similar parameter setting. Additionally, we use the non-equilibrium Green's function (NEGF) method to track the evolution of propagating edge heat current contributions that can flip their orientation. This further confirms the presence of the sign change behavior. Finally, we propose candidate square-octagon materials where our theory can be tested.

This paper is organized as follows. In Sec.~\ref{secII} we define the model (Sec.~\ref{secIIA}). Next, we introduce the spin wave and Schwinger boson theory formulation (Sec.~\ref{secIIB}), Chern number concept and thermal Hall conductance definition (Sec.~\ref{secIIC}), and the NEGF formalism (Sec.~\ref{secIID}). In Sec.~\ref{secIII} we present our results on topological energy bands (Sec.~\ref{secIIIA}), topological phase diagram (Sec.~\ref{secIIIB}), thermal Hall effect (Sec.~\ref{secIIIC}), the Einstein-de Haas effect (Sec.~\ref{secIIID}), and finally we discuss the materials significance of our results (Sec.~\ref{secIIIE}). In Sec. IV we discuss and conclude our findings. In the subsequent appendices we present the equations for spin-wave Hamiltonians for the honeycomb and the kagome lattices (Appendix~\ref{appdA}), the expressions for Schwinger boson mean field theory equations (Appendix~\ref{appdB}), the magnetic and topological phase diagrams (Appendix~\ref{appdC}), edge state geometry of a strip sample (Appendix~\ref{appdD}), thermal Hall conductance and angular momentum expressions (Appendix~\ref{appdE}), the NEGF equations (Appendix~\ref{appdF}), and the realistic materials parameters (Appendix~\ref{appdG}).

\section{Model and methods}\label{secII}
\subsection{Square-octagon spin model}\label{secIIA}

We consider a spin-1/2 Heisenberg system on a square-octagon lattice, see Fig.~\ref{fig:fig1}. As shown, the geometry consists of four two-dimensional square sub-lattices obtained by replacing each site of a square lattice by a tilted $45^\circ$ square plaquette. The side of every octagon and the square is set to $(\sqrt{2}-1)a$, where $a$ is the edge length of the square unit cells, see red dashed square in the figure. The model Hamiltonian contains nearest-neighbor (nn) ferromagnetic (FM) Heisenberg exchange interaction, the nn and next-nearest-neighbor (nnn) DM interaction terms (see Fig.~\ref{fig:fig1}), and an external Zeeman magnetic field term. The total Hamiltonian is given by
\begin{equation}
\begin{split}
\mathcal{H} &= -\sum_{\langle mn \rangle} J_{mn} \bm{S}_{m} \cdot \bm{S}_{n} \\
&+\sum_{\langle{mn}\rangle, \langle\langle mn \rangle\rangle}\bm{D}_{mn} \cdot (\bm{S}_{m} \times \bm{S}_{n}) - h \sum_{m} S_{m}^z,
\end{split}
\end{equation}
where $\bm{S}_m$ denotes the spin angular momentum at site $m$, $h=\mu_{B} g H$ is the magnetic field interaction, $g$ is the $g$-factor, $\mu_B$ is the Bohr magneton, and $H$ is the external magnetic field. The ferromagnetic exchange coupling is chosen to be positive, $J_{mn}>0$. We set $J_1$ and $J_2$ equal to unity and take the spin to be $S=1/2$. The DM interaction $D_{mn}$ is over the nn and the nnn neighbors. As shown in Fig.~\ref{fig:fig1}(b), the DM interaction is tuned over a range of parameters to study the EdH effect in the square-octagon model. The nonzero DM terms combine with the exchange interaction to generate a complex phase factor. Hence, the presence of DM coupling imparts a non-trivial topological character to the bands.

\subsection{Spin wave and Schwinger boson theory}\label{secIIB}

\begin{figure*}[t]
	\centering
	\includegraphics[width=0.98\linewidth]{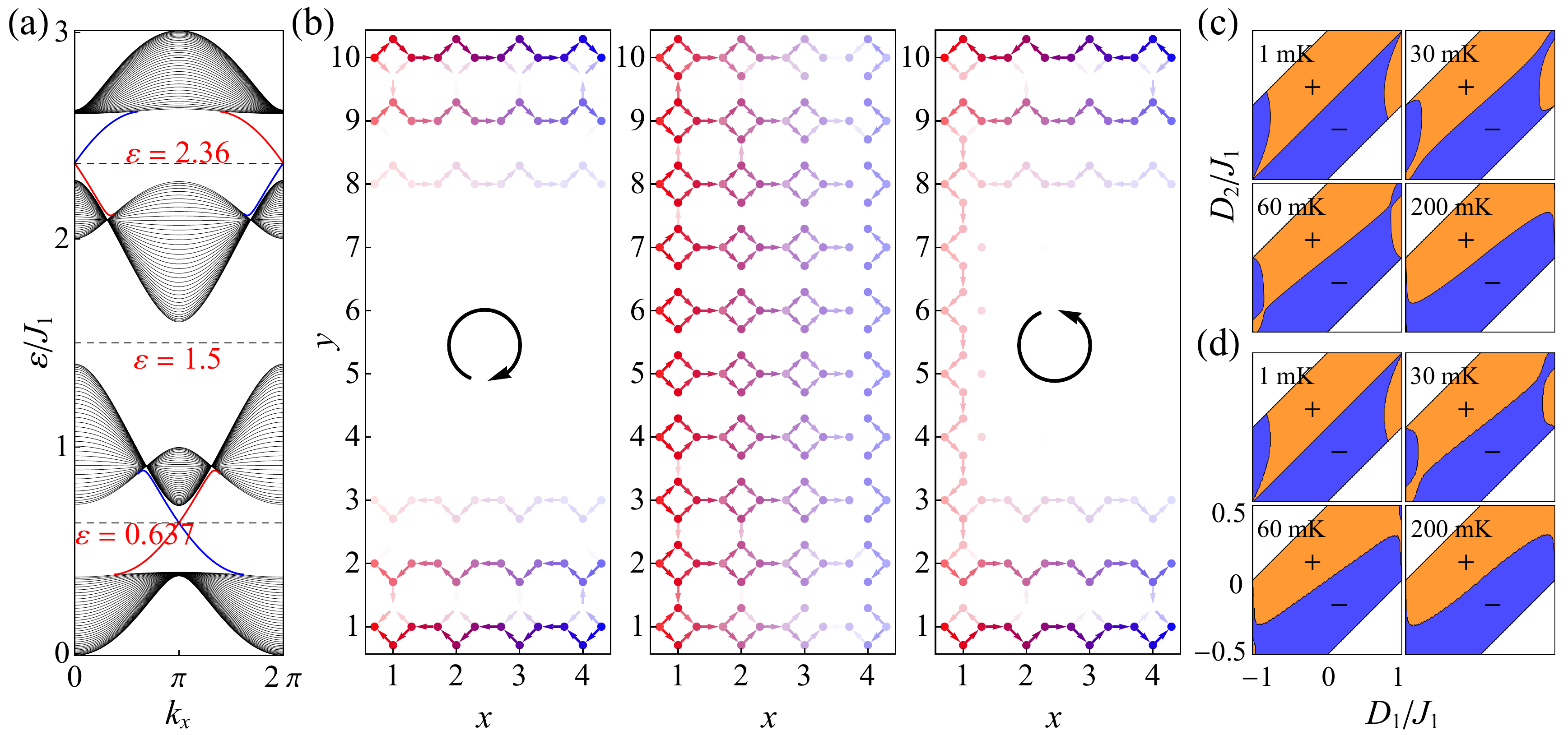}
	\caption{\label{fig:fig2} Heat current orientation reversal with temperature in a square-octagon lattice.~(a)~Energy band diagram corresponding to topological edge states for the $A_{+0}$ phase ($D_{1,2}/J_1=0.6,0$ with $J_1 = 1 K$), see Fig.~1. The dashed lines indicate the choice of energy levels for the non-equilibrium green function (NEGF) calculation.~(b)~NEGF heat current plots displaying thermal Hall current direction reversal. Energy selection of the left, middle, right plots are $0.637J_1$, $1.5J_1$, and $2.36J_1$ respectively. The temperature of the left and right leads are set to $T_L=2.5J_1$ and $T_R=1.5J_1$. The opacity of color represents the magnitude of local magnon density of states.~(c)~Edge angular momentum evolution with temperature. Positive and negative regions of circulation are colored orange and blue, respectively.~(d)~Corresponding $\kappa_{xy}$ temperature evolution. Panels {(c)} and {(d)} track the sign change evolution at four different temperatures ranging from 1 to 200 mK.}
\end{figure*}

In order to study topological magnon excitations, we consider a linear spin-wave theory approach in the low temperature regime where magnon excitations are well-defined. The spin Hamiltonian is transformed into a magnon Hamiltonian following the usual Holstein-Primakoff transformations: $S_{m}^{+} = \left(2 S - b_{m}^{\dag} b_{m}^{}\right)^{1/2} b_m$, $S_{m}^{-} = b_{m}^{\dag} \left(2 S - b_{m}^{\dag} b_{m}^{}\right)^{1/2}$, $S_{m}^{z} = S-b_{m}^{\dag} b_{m}^{}$. Only bilinear terms are retained. Higher order interaction effects will not change the main conclusions of our article. The Hamiltonian has the form
\begin{equation}
\begin{split}
\mathcal{H} &=-\sum_{mn} \left[\left(J_{mn} + i \nu_{mn} D_{mn}\right) S b_{m}^{\dagger} b_{n}^{} + \text{H.c.}\right]\\
&+ h \sum_{m} b_{m}^{\dag} b_{m}^{} + E_{0},
\end{split}
\end{equation}

where $E_0$ is ground state energy. $\nu_{mn}=-\nu_{nm}$ designates the direction of DM interaction. Subsequently, energy bands are obtained via diagonalization of the Fourier transformed Hamiltonian matrix. The edge states were calculated for a strip sample in open boundary condition. In Fig.~\ref{fig:fig2}, we considered $W=120$, which is large enough to fix the edge states~\cite{hatsugai1993edge}. Different edge conditions will give rise to different edge states. But, the topology features remain unchanged. Our choice of edge boundary is shown in Fig.~\ref{appdfig4}.

While the magnon picture is effective at low temperatures and offers explanation of experimental data~\cite{onose2010observation}, it fails at higher temperatures where thermal fluctuations cannot be neglected. The bilinear magnon approximation $\sqrt{2 S-b^{\dag}b}\rightarrow \sqrt{2 S}$ becomes invalid. We identify the disordering temperature as the value where the $\langle b^\dag b \rangle$ reaches $2 S$. Note, we are not concerned with the nature of the transition. Rather our focus is on understanding how either the ordered or the disordered quantum magnetic states behave. Consequently, we utilize the Schwinger boson mean-field theory (SBMFT)~\cite{sarker1989bosonic} formalism for the thermally disordered quantum paramagnetic regime~\cite{kim2016realization}. The Schwinger boson approach is capable of describing experimental results~\cite{lee2015thermal} and can additionally give rise to the spin Nernst effect~\cite{kim2016realization}. In contrast to spin-wave theory, SBMFT considers the following transformations --- $\bm{S}_m= b_{ms}^{\dag}(\bm{\sigma})_{st}b_{mt}^{}/2$, where $\sigma$ are Pauli matrices, $s$ and $t$ are spin indices. The bosonic commutation relations $[b_{ms}^{},b_{nt}^{\dag}]=\delta_{mn}\delta_{st}$ keep the spin algebra invariant. As an approximation, ten mean-fields $\lambda_i$ are defined to describe the ground state. The spinon excitation was taken as a perturbation by using the Hartree-Fock decoupling scheme where $AB\mapsto \langle A\rangle B+A\langle B\rangle-\langle A\rangle\langle B\rangle$. This reduces the quartic interaction terms of the Hamiltonian to bilinear contributions. The value of the mean-fields were determined by minimizing the free energy $F=F_0(\lambda)+k_B T\sum_{\bm{k}\mu s}\ln \left(1-\exp\left(-E_{\bm{k}\mu s}/k_B T\right)\right)$. Ten self-consistency equations $\partial F/\partial \lambda_i=0$ ($i = 1, \dots, 10$) were numerically solved to obtain the results. Please see Appendices~\ref{Appd:A} and \ref{Appd:B} for further details on spin-wave and Schwinger boson theory. The spinon results are not reported in the main article. The nonzero EdH response of spinons is discussed in Appendix~\ref{Appd:E}.

\subsection{Chern number and thermal Hall conductance}\label{secIIC}

The thermal Hall effect can be considered as the bosonic version of the integer quantum Hall effect. Thus some properties of topological insulators can also be found in our system, such as topological numbers, and conducting edge states. The topological invariant of our topological magnon state is the Chern number~\cite{hasan2010colloquium}, which is the integral of the Berry curvature over the Brillouin zone. The Berry curvature of a band $\lambda$ is given by
\begin{equation}\label{BerryCurvature}
\Omega_{\lambda \bm{k}}=i \sum_{\mu\not=\lambda} \frac{\left[\braket{\lambda|\nabla_{\bm{k}} H(\bm{k})|\mu} \times \braket{\mu|\nabla_{\bm{k}} H(\bm{k})|\lambda}\right]_z}{(E_{\lambda}-E_{\mu})^2}.
\end{equation}
\noindent The Chern number is given by the integral 
\begin{equation}\label{ChernNumber}
C_{\lambda}=\frac{1}{2\pi}\int_{\text{BZ}} \mathrm{d^2}k \Omega_{\lambda \bm{k}}.
\end{equation}
The boundaries of the topological phase diagram was determined based on tracking gap closures and opening~\cite{hatsugai1993edge,thouless1982quantized}.

The electronic Hall conductance, originating from the Kubo formula, is the integral of the Berry curvature with an appropriate weight, which relates the anomalous velocity of the magnons, multiplied by the Fermi distribution function over all energies~\cite{thouless1982quantized,kohmoto1985topological}. With this description, the topological phase is still included within the framework of Bloch's wave function and band theory, that is, topological band theory. Similarly, the thermal Hall conductance can be expressed as a weighted summation of the Berry curvature, according to linear response theory as~\cite{matsumoto2011rotational,matsumoto2011theoretical}
\begin{equation}\label{ThermalHallConductance}
\kappa_{xy}=-\frac{k_B^2 T}{\hbar a}\frac{1}{N}\sum_{\lambda\bm{k}} c_2(n_B(\varepsilon_{\lambda\bm{k}})) \Omega_{\lambda \bm{k}},
\end{equation}
where the weight function is given by $c_2(x)=(1+x)(\ln (1+1/x))^2-(\ln x)^2-2\text{Li}_2 (-x)$, where $\text{Li}_s (z)=\sum_{n=1}^{\infty}z^n/n^s$ is the polylogarithm function. $c_2$ is a decreasing function similar to the Bose distribution. The lattice constant is chosen as $a=0.1\text{nm}$ in this article. Further details on thermal Hall conductance can be found in Appendix~\ref{Appd:E}.

\subsection{Non-equilibrium Green's function (NEGF)}\label{secIID}

We perform non-equilibrium Green's function (NEGF) calculations on the square-octagon lattice model to solidify our proposal for the EdH effect. For this purpose, the sample was divided into three parts --- the left lead, the central part, and the right lead. The lead temperatures were kept at unequal $T_L$ (left) and $T_R$ (right) values to ensure a thermally non-equilibrium state. The magnon transport of the central part is what we are concerned with~\cite{zhang2013topological}. The total Hamiltonian can be written as
\begin{equation}\label{NEGFHamiltonian}
\mathcal{H}=\sum_{\alpha, \beta=L, C, R} b_{\alpha}^{\dag} H_{\alpha \beta} b_{\beta}.
\end{equation}
The Hamiltonian matrix is then used to compute the self energy, the lesser, the retarded, and the advanced Green's function of the central part. Furthermore, quantities such as the local density of magnons and the local current were determined from the Green's functions as
\begin{equation}\label{NEGFDensity}
\rho_{m} = \frac{i\hbar (G_{CC}^<)_{mm}}{\pi a},
\end{equation}
\begin{equation}\label{NEGFCurrent}
j_{mn} = \frac{\varepsilon}{2\pi} \text{Re}[(G_{CC}^<)_{mn}H_{nm}-(G_{CC}^<)_{nm}H_{mn}].
\end{equation}

The two leads and the central part are supposed to be large enough to eliminate the influence of boundaries. But, in practice, for a schematic calculation a $4\times 10$ lattice site discretization suffices to demonstrate the basic EdH effect, see Fig.~\ref{fig:fig2}. For further calculation details, please refer to Appendix~\ref{Appd:F}.

\section{Results}\label{secIII}
\subsection{Topological energy bands}\label{secIIIA}

The topological identity of energy bands can be characterized by nonzero Chern numbers of bulk bands or edge states in a finite-size system. Within our model, considering a strip sample, we draw edge states as shown in Fig.~\ref{fig:fig2}(a). In the figure, the dense overlapped states are bulk states, which are also the projection of energy bands in the $k_y$ direction. The states lying in the gaps and connecting two bulk bands are edge states. 
In the absence of DM interaction, the first (second) band touches the third (fourth) band at the $\text{M}(\pi,\pi)$ ($\Gamma(0,0)$) point. The intermediate energy bands are flat at those high symmetry points. With a nonzero DM term, we can open gaps at those degenerate points to turn the system into a magnon insulator~\cite{owerre2016first}. The gap expressions are $\pm 2(D_1+2D_2)S$ or $ \pm 2|D_1-2D_2|S$ at the $\Gamma$ and the $\text{M}$ points, respectively. See Appendix~\ref{Appd:C} for more detail. Note, Chern numbers are well-defined in band insulators, only.

\subsection{Phase diagram}\label{secIIIB}

The topological and magnetic phase diagram for a set of model parameters is shown in Fig.~\ref{fig:fig1}. The FM and non-FM regions are highlighted in the $(D_2/J_1,D_1/J_1)$ plane. The zero temperature phase diagram was constructed via energy comparison between non-ferromagnetic versus ferromagnetic states (which have the lowest energy). The topological phases, characterized by Chern numbers~\cite{hasan2010colloquium}, are shown in Fig.~\ref{fig:fig1}. The bulk-boundary correspondence can be seen in Fig.~\ref{fig:fig2}. According to Fig.~\ref{fig:fig1}, the Chern number combination is $\{1,-1,-1,1\}$. Since the winding numbers are $w_1=-w_3=1$, $w_2=0$, there is only one loop of edge state in the first and the third gap with opposite orientation. Hence, there is no edge state in the second gap. The group velocity of the edge states show a chirally moving magnon current localized at the edges, whose direction is consistent with the total winding number in that gap. A thorough phase diagram calculation is reported in the Fig.~\ref{appdfig2}.

\subsection{Thermal Hall effect}\label{secIIIC}

We studied the topological magnon Hall effect using the NEGF formalism, see Fig.~\ref{fig:fig2}. Calculation details are provided in Appendix~\ref{Appd:F}. A temperature gradient was introduced to a finite sample to study the magnon heat current along the $x$ direction. The generation of Berry curvature (due to the presence of DM interaction) leads to a current deflection in the $y$ direction. This results in chiral edge current transport. The chiral current is localized. For a selected energy level in the gap, the orientation of the chiral current corresponds well with the sign of the winding number for the edge states as shown in Fig.~\ref{fig:fig2}(a). The edge condition and calculation details can be found in Appendix~\ref{Appd:D}. With a temperature gradient along the x-direction, there is a net thermal Hall conductance between the top and the bottom edge~\cite{matsumoto2011theoretical}. There is also a net magnon current transport from the hotter to the colder part of the sample~\cite{zhang2013topological}. Within our parameter setting, each sub panel of Fig.~\ref{fig:fig2}(b) represents the $\kappa_{xy}$ current corresponding to the dashed energy levels of $0.637J_1$, $1.5J_1$, and $2.36J_1$ in the gap, respectively. As the figure shows, there is a net nonzero current for the left and the right sub panels, but there is no net edge state current in the second gap.

Theoretical~\cite{Mook2014Edge,mook2014magnon,lee2015thermal,cao2015magnon} and experimental~\cite{hirschberger2015thermal} studies have shown that thermal Hall conductance can change sign as temperature or magnetic field is varied. In Fig.~\ref{fig:fig2}(c) and \ref{fig:fig2}(d), we show the sign evolution of angular momentum together with the variation of thermal Hall conductance with temperature. See Appendix~\ref{Appd:E} for details on how these plots were obtained. The shape change of the boundary implies that previously identified positive zones have now been converted into negative areas (the blue color comes orange). We can see from Fig.~\ref{fig:fig2}(c) and \ref{fig:fig2}(d) that there is a pinch point at around 60 mK where the transition happens. Currently, several competing theoretical proposals have attempted to explain the thermal Hall sign change behavior. The theoretical constructs range from the high temperature limiting value of thermal conductance to the Chern number of bands~\cite{mook2014magnon}, propagation direction of edge modes~\cite{Mook2014Edge}, non-uniform Berry curvature over the Brillouin zone~\cite{lee2015thermal}, and sensitivity of weight function in the thermal Hall formula~\cite{lee2015thermal}. The importance of the weight function is manifested in the sign change behavior of Nernst conductivity~\cite{cheng2016spin}. Irrespective of the conceptual reason for sign variation, we show that a measurable EdH mechanism can be observed in a host of frustrated lattices. Note, that thermal Hall effect is an important probe of DM interaction. Based on our studies we can infer that when the DM interaction is large enough the EdH effect can then be observable.

\subsection{Einstein-de Haas effect}\label{secIIID}

The EdH effect is the ultimate macroscopic manifestation of a very subtle microscopic exchange of angular momentum originating from the electron spin (an inherently quantum object). The quantum to classical transfer is succinctly summed up in the equation
\begin{equation}
\Delta L_{\text{sample}}+\Delta L_{\text{inner}}=0.
\end{equation}
Magnons carry angular momentum in units of $\hbar$. However, in a topological magnon insulator the magnon carries an additional angular momentum originating from Berry curvature~\cite{matsumoto2011rotational}.

\begin{figure}[h]
	\centering
	\includegraphics[width=0.98\linewidth]{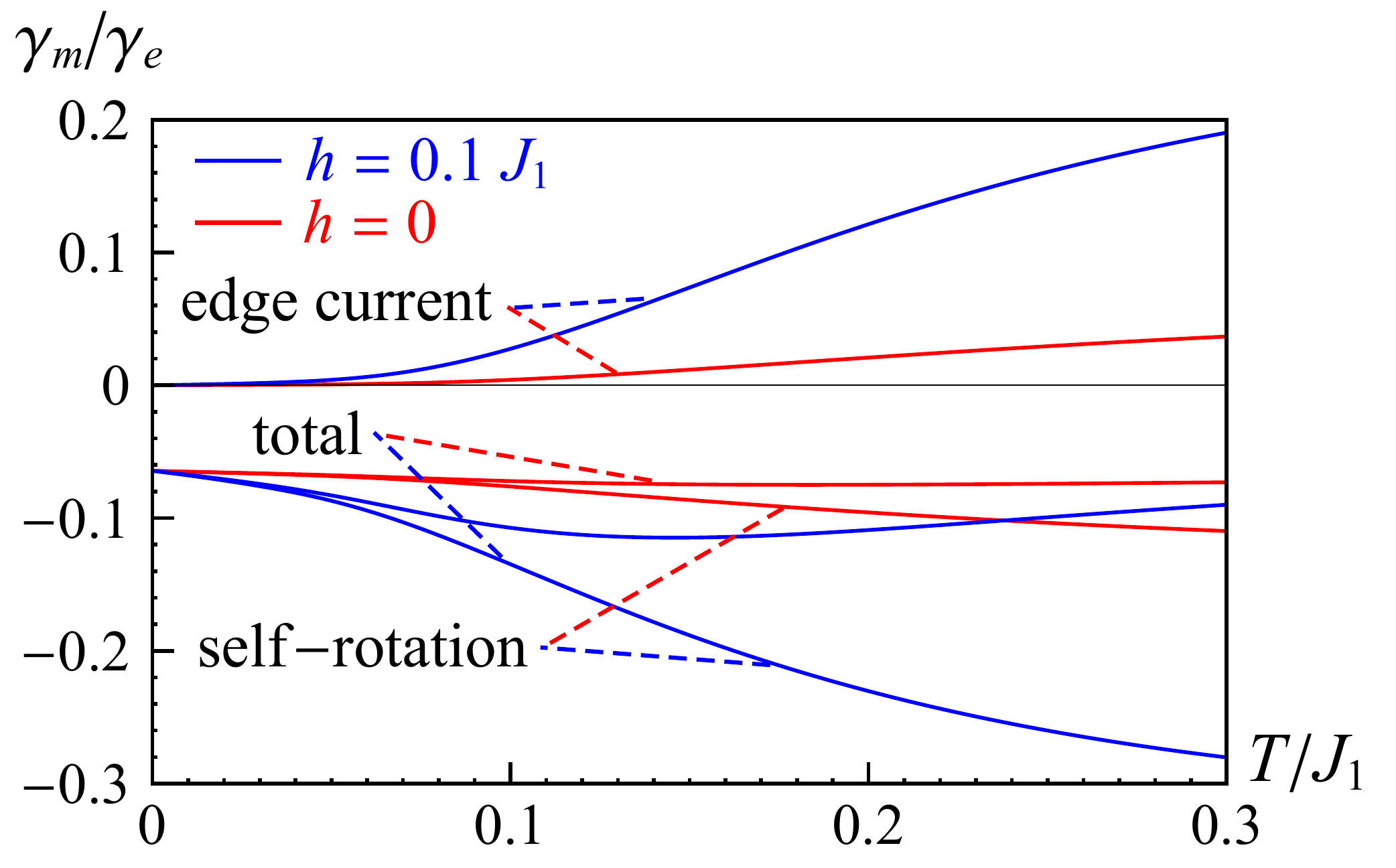}
	\caption{\label{fig:fig3} Temperature dependence of the topological gyromagnetic ratio $\gamma_m$ in a square-octagon lattice.~The gyromagnetic ratio contribution, compared to the electronic value, for individual topological edge current, self-rotation of the magnon wave packet, and the total angular momentum contribution to $\gamma_m$ are shown, both in zero and nonzero external field $h$. $J_1$ is the nn ferromagnetic interaction. Parameters choices were $D_{1,2}/J_1=0.6, 0.1$.}
\end{figure}
\begin{figure*}[t]
	\centering
	\includegraphics[width=1.0\linewidth]{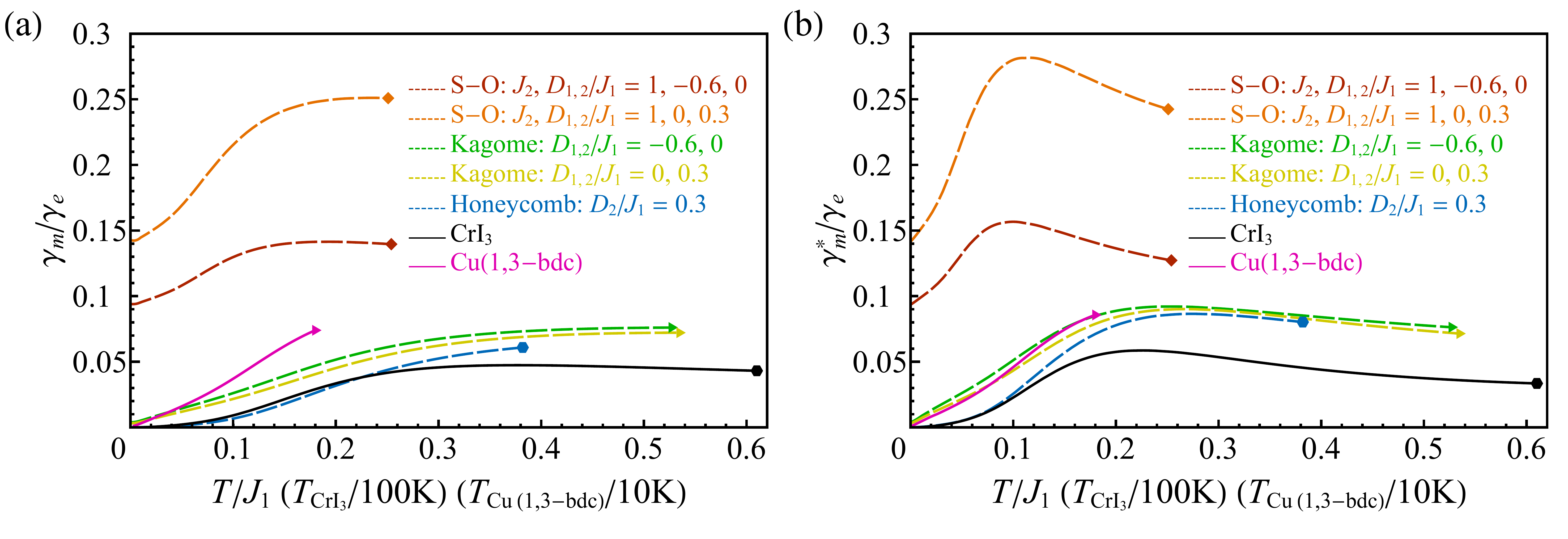}
	\caption{\label{fig:fig4}Comparison of the Einstein de-Haas effect response in square-octagon, kagome, honeycomb lattice, $\text{Cu(1,3-bdc)}$, and $\text{CrI}_3$. Plot markers at the end of the line plot indicate the termination point beyond which our theory ceases to be valid.~(a)~The bare topological to electronic gyromagnetic ratio variation with temperature is shown.~(b)~The differential gyromagnetic ratio response. In both panels the plots are truncated at the thermal disordering temperature. The saturation temperature of magnetization is given by $T(\langle b^\dag b \rangle = 2\text{S})$. For the realistic materials the Curie temperature $T_c$ is measured by experiments.}
\end{figure*}
The EdH effect of topological magnons occurs when the system is exposed to a rapidly changing magnetic field or temperature. The sudden change causes the magnetization of the system to react. Thus, an angular momentum deviation, related to the magnetization change by the gyromagnetic ratio $\gamma_e$, is inflicted. Concurrently, topological magnons, will experience an additional angular momentum change resulting from the Berry curvature. This consists of the self rotation and the edge current of magnons. So the key to estimating the magnitude of the EdH effect of a topological magnon insulator is the gyromagnetic ratio of magnons. We can define the gyromagnetic ratio of magnons as the angular momentum divided by the magnetic moment of magnons, which is related to the magnetization change of the system. The magnetic moment of the magnons is expressed as
\begin{equation}
\Delta M=-\frac{\hbar}{\gamma_e N} \sum_{n\bm{k}} n_B(\varepsilon_{n\bm{k}}),
\end{equation}
where the gyromagnetic ratio of our spin system is $\gamma_e=2m_e/(ge)$; $g$ is the Lande factor, $e$ and $m_e$ are the charge and mass of the electron, respectively. Within the low temperature approximation, the mass of the magnon can be approximated as the effective mass at the bottom of the first band, an approximation which is also valid at high temperature due to the Bose-Einstein distribution.
Thus, the gyromagnetic ratio of topological magnons is given by
\begin{equation}
\gamma_m=\frac{L_{tot}}{\Delta M},
\end{equation}
where $L_{tot}$ represents the total sum of the edge and self-rotation angular momentum. This definition is motivated purely from an intuitive and fundamental understanding of the gyromagnetic ratio~\cite{PhysRevSeriesI.26.248}. The result of our calculation is shown in Fig.~\ref{fig:fig3}. We find two competing angular momentum components which oppose each other. The self-rotation part is larger in magnitude, but with an opposing sign. The total gyromagnetic contribution has a nonzero value in the zero temperature limit. Note, the additional amplitude of the EdH effect of topological magnons, compared to the pure two dimensional ferromagnetic system, is appreciable as long as the DM interaction is not too small. To further analyze the physical content of Fig.~\ref{fig:fig3}, we compare and contrast the topological gyromagnetic ratio with respect to the kagome and the honeycomb lattice systems.

In Fig.~\ref{fig:fig4} we show the results of our $\gamma_m/\gamma_e$ calculation. We notice that in the very low temperature limit (almost zero), the square-octagon lattice has a nonzero response. This sets it apart from the other usual lattice candidates such as kagome and honeycomb, that have been experimentally explored till now. The zero-temperature EdH response of these lattices is at least two to four orders of magnitude smaller. In Fig.~\ref{fig:fig4}(a) we show the temperature variation of the topological gyromagnetic ratio compared to the electronic value. The curve is truncated at a point where the magnetic ordering begins to enter a high temperature disordered paramagnetic phase. The disordering temperature for the various lattices studied in this article are listed in the Table.~\ref{appdtab1}. While initially the system is sensitive to temperature changes, at higher temperatures the response hits a plateau. This could potentially be due to magnon excitations not being well-defined anymore as we encroach the thermal disordering temperature. Although, we have demonstrated the existence of the EdH effect in only these three lattices, our formalism, analysis approach, and eventual conclusions will hold for a wider variety of frustrated ferromagnetic systems.

In contrast to the topological gyromagnetic ratio response, the EdH caused by electrons is determined by the electron itself. Considering this, we define a differential gyromagnetic ratio response $\gamma^{*}_m$ of the topological magnons as
\begin{equation}
\gamma^{*}_m=\left(\frac{\partial L_{tot}/\partial T}{\partial \Delta M/ \partial T}\right)_h.
\end{equation}
For electrons $\gamma_e^{*}$ is equal to $\gamma_e$. However, the situation is entirely different for topological magnons which are excitation quasi-particles. For these excitation modes the gyromagnetic ratio is renormalized from the bare $\gamma_m$ response. Each magnon mode has its own gyromagnetic response. Additionally, the magnons can be excited or annihilated. Thus, experimentally the gyromagnetic ratio response cannot simply be measured from $\gamma_m$ itself, but from a response to a temperature change. Evaluating the above equation shows that the system does have a maximal response as seen in Fig.~\ref{fig:fig4}(b), before dipping off. A magnetic field can enhance the EdH, just as in Fig.~\ref{fig:fig4}(b). Thus, there is an optimal temperature at which the magnon insulator will have the best possible response. Thus, from an experimental point of view, this is the optimal temperature zone in which our theory can be tested. For the square-octagon lattice this value is around $T/J_1 = 0.1$. For the kagome and the honeycomb the ratio is about 0.24.

\subsection{Materials significance}\label{secIIIE}

Our discussion up to this point has highlighted the presence of sizable EdH effect in three kinds of two-dimensional lattices. The results show that the square-octagon lattice exhibits the largest effect, under similar parameter settings. Till date the existence of topological magnon excitations have not been confirmed experimentally in the square-octagon lattice. Thus, it is natural to ask (and vital for experimental verification) --- is the EdH effect large enough in existing materials which can harbor topological magnons? We study a couple of realistic two-dimensional materials --- $\text{CrI}_3$ \cite{Chen2018TopologicalSE} and $\text{Cu(1,3-bdc)}$ \cite{hirschberger2015thermal,chisnell2015topological,nytko2008structurally}. The spin Hamiltonians and their parameters are listed in Appendix~\ref{Appd:G}. These materials have some notable distinctions. Whereas $\text{CrI}_3$ is an $S=3/2$ honeycomb monolayer ferromagnet, $\text{Cu(1,3-bdc)}$ is an $S=1/2$ kagome ferromagnet. In $\text{CrI}_3$ the ferromagnetic coupling is large enough to result in ferromagnetism, but in $\text{Cu(1,3-bdc)}$ a magnetic field ($\mu_0 H > 0.05\text{T}$) is necessary to polarize all spins due to the weak interlayer antiferromagnetic coupling. \cite{chisnell2015topological}. The Curie temperature of $\text{CrI}_3$ ($\approx 61 \text{K}$) is much higher than $\text{Cu(1,3-bdc)}$ ($\approx 1.8 \text{K}$), which may affect the experimental feasibility. The EdH response of these materials are shown in Fig.~\ref{fig:fig4}. The curves are truncated at the Curie temperature, which is much smaller than the disordering temperature given by the bilinear magnon approximation ($162\text{K}$ for $\text{CrI}_3$ and $16.7 \text{K}$ for $\text{Cu(1,3-bdc)}$). Similar to the model parameter calculation we notice a nonzero EdH response in realistic materials. The EdH effect of the kagome lattice is larger than the honeycomb lattice, even though $D_1/J_1=0.15$ of $\text{Cu(1,3-bdc)}$ is not twice the value $D_2/J_1=0.154$ of $\text{CrI}_3$. One should notice that their temperature scales are quite different. However, the EdH effect in these materials is still observable.

\section{Discussion}\label{secIV}

In summary, we predict the existence and detection of Einstein-de Haas effect of topological magnons in the square-octagon, kagome, and honeycomb lattices. The influence of a nonzero Berry curvature originating from DM interactions and its underlying topological identity is preserved even though the lattice structure changes. Thus, similar conclusions, not qualitatively differently, can be obtained from other lattices which can support a nonzero Berry curvature. Our conclusions should also hold for other recently studied systems such as the kagome lattice~\cite{Mook2014Edge,seshadri2018topological} and the Lieb lattice~\cite{cao2015magnon}, or the star lattice~\cite{owerre2016magnon}. It is also interesting to consider a model with two kinds of DM interaction, competing with each other on the square-octagon lattice. More precisely, the system should possess a broken spin up-down symmetry as in a ferro- or ferri- magnetic system. In antiferromagnets, the DM interaction can only result in the spin Nernst effect~\cite{cheng2016spin}.

We have shown that the EdH effect in $\text{Cu(1,3-bdc)}$ and $\text{CrI}_3$ is observable. But, we would like to encourage experimentalists to find an appropriate square-octagon material, since it has the largest EdH effect, especially in low temperature regime, where the magnon description is more effective. At present, it is challenging to find an appropriate material that has large enough DM interactions. Here we discuss some possible realistic materials where our theory could be tested. Recent studies have predicted two-dimensional square-octagon compounds in the C or the group-V materials~\cite{sheng2012octagraphene,guan2014tiling,zhang2015two,kou2015tetragonal,ersan2016stable}. Monolayer or multi-layer (V-III bonded) Haeckelite compounds are other potential examples~\cite{camacho2015gan,shahrokhi2017new,gurbuz2017single}. The square-octagon $\text{XY}_2$ compound, with two sublattices, has been recently synthesized, too~\cite{li2014gapless}. If magnetic dopants could be introduced into the X-sublattice, then the bilayer structure is another potential candidate. Besides the above examples, quasi two-dimensional materials can possess intrinsic ferro- or antiferromagnetism in a nearly square-octagon structure, also~\cite{ma2013competing,calta2013quaternary}. In these compounds, Fe or Mn transition metal ions form an interpenetrating square-octagon layer with the rare earth element residing in an almost square-octagon geometry. If we remove the central atom of the rare-earth sublattice to introduce a vacancy, or dope all the rare earths by nonmagnetic atoms, a square-octagon structure as proposed in our model could be achieved.

In Stewart's original experiment~\cite{Stewart1918}, he observed the EdH effect in iron and nickel. In his set up, the sample was suspended inside a solenoid by a quartz fiber. A changing current in the solenoid circuit caused an angular momentum impulse. This generated a mechanical rotation of the suspended system. The rotation angle was measured as the deflection of a spot of light that reflected from a mirror attached to the freely suspended system. The  deflection was proportional to the angular momentum change originating from the EdH effect. The rotational motion of the system was dictated both by material properties and the amount of the transient angular momentum change. 

\begin{figure}[h]
	\centering
	\includegraphics[width=0.98\linewidth]{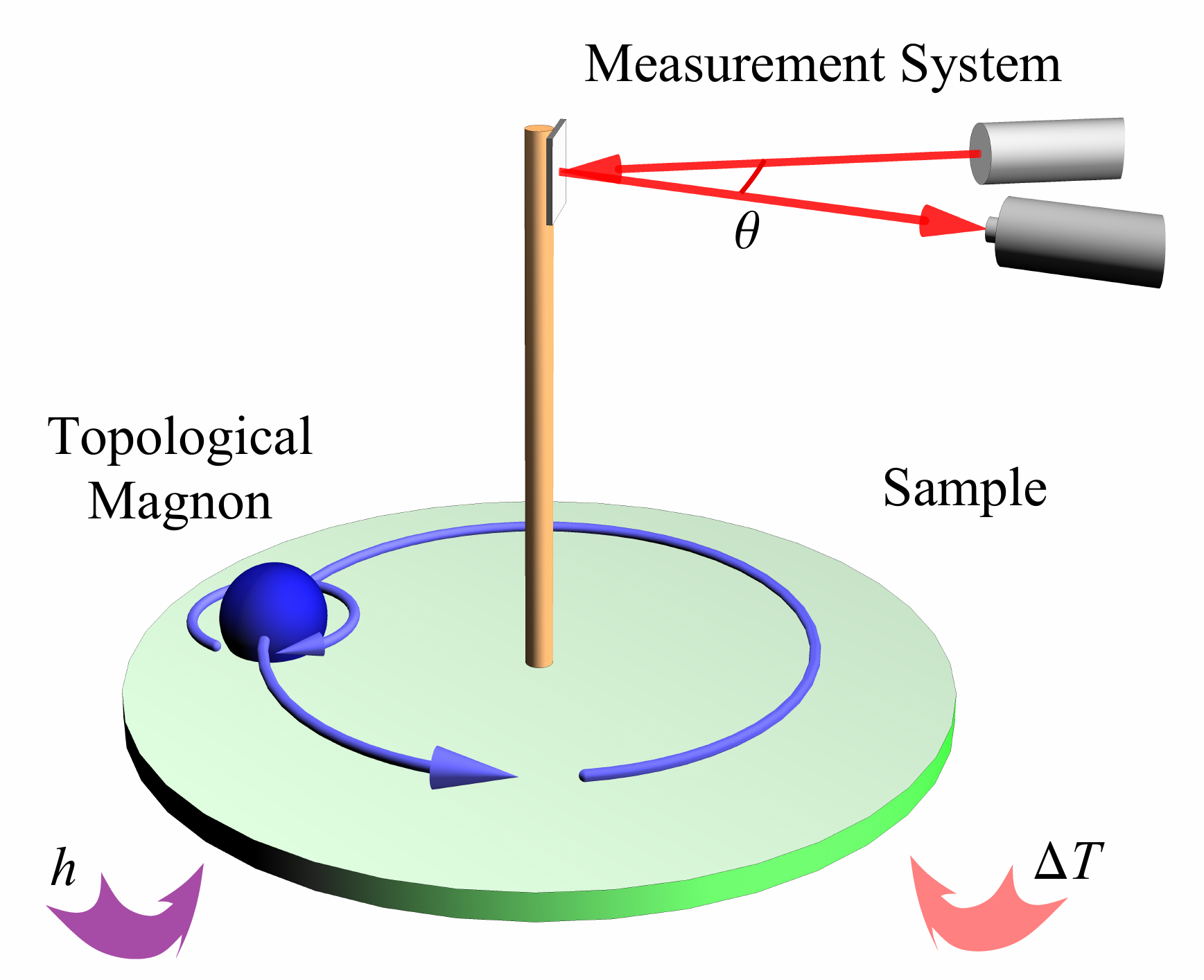}
	\caption{\label{fig:fig5} Proposed experimental set up of the Einstein-de Haas effect of topological magnons. The disk shaped sample is exposed to an external heat bath with a temperature gradient and also to an external field $h$, if required. The mirror mounted on the axle of the disk will rotate and deflect the incident light. The deflection angle is proportional to the Einstein de-Haas effect.}
\end{figure}

To test our theoretical prediction, we propose an experimental set-up similar to that of Stewart as shown in Fig.~\ref{fig:fig5}. The difference is that our system is two-dimensional. Thus, the sample is disk shaped. Immediately below the sample there is a heater or an electric plug to adjust the temperature or magnetic field. Note, a strong magnetic field can also polarize the ground state to a ferromagnetic phase. The sample is connected to a suspension wire or equipment suitable enough to provide the torque. The rotational force can be measured by a detecting device such as a mirror attached to the wire with a light beam and a light sensor, or some form of atomic force detector. In an actual experiment, the temperature needs to be appropriately adjusted so that it is neither too high nor too low, see Fig.~\ref{fig:fig4}(b) for guidance. Note, in Stewart's paper, his differential equation response was modeled as 
a damped harmonic oscillator characterized by a damping coefficient and a moment of inertia. The initial angular momentum impulse, which appeared as an initial condition, was determined by measuring the first amplitude of oscillation. If the EdH of magnons is ultrafast, which implies the time scale of the inner angular momentum transferred to the whole sample is much shorter than those of the relaxation time in the experiment, then our model is applicable. If not, we need to consider a new microscopic mechanism of spin-to-rotation transfer.

The observed experimental results could differ from theoretical prediction due to several reasons. First, the bilinear magnon approximation is too crude. Thus one has to include interaction terms by mean field theory or Green's function method to get the many body corrections. Second, contributions from other kinds of topological effects such as topological phonons could arise. Phonons can also contribute to thermal Hall effect and the EdH effect \cite{zhang2014angular, zhang2016berry}, with some topological terms included in Hamiltonian. However, this may not be a concern since systems in which topological magnons exist already dominate the thermal Hall effect behavior.

\appendix
\section{\label{Appd:A}Spin-wave Hamiltonian}\label{appdA}
\begin{figure}[t]
	\centering
	\includegraphics[width=\linewidth]{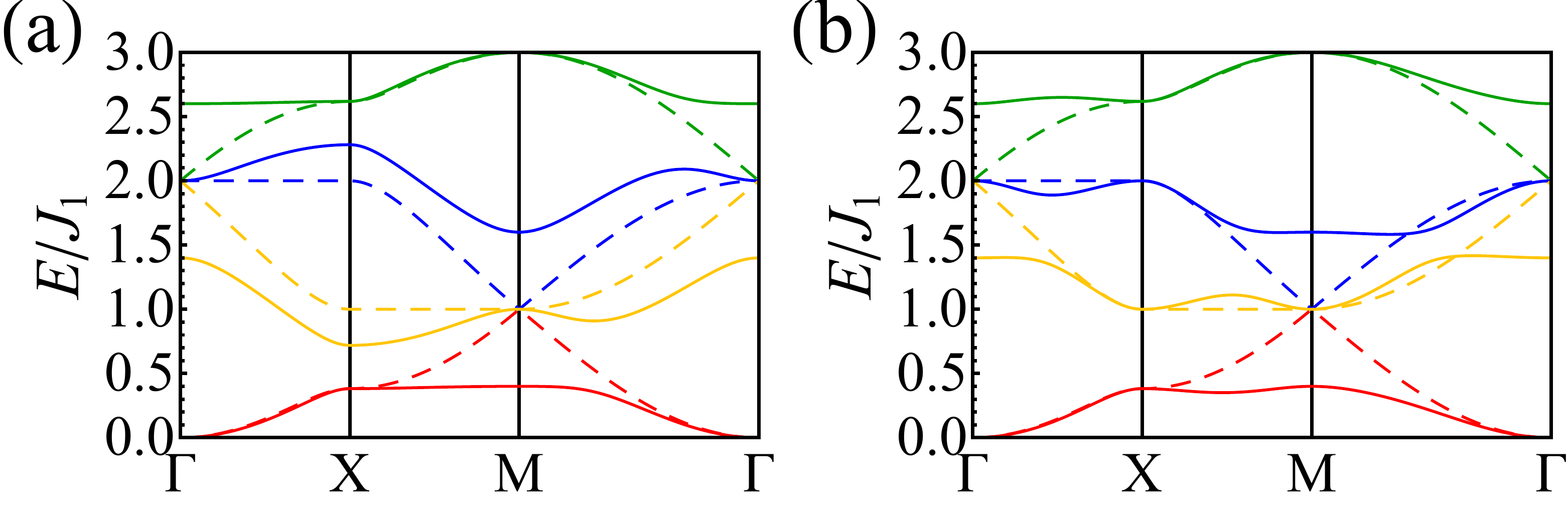}
	\caption{\label{appdfig1} Magnon bands of square-octagon lattice along $\Gamma(0,0)-\rm{X}(\pi,0)-\rm{M}(\pi,\pi)-\Gamma(0,0)$ including DM interactions. The dashed lines are bands without DM interactions, the solid lines are bands with nonzero DM interactions. The parameters are set as: (a)~ $S=1/2$, $J_1=J_2=1.0$, $D_1=0.6 J_1$, $D_2=0$. (b)~ $S=1/2$, $J_1=J_2=1.0$, $D_1=0$, $D_2=0.3 J_1$.}
\end{figure}

The spin-wave Hamiltonian has an exact ferromagnetic ground state. The magnon system can be studied using the Holstein-Primakoff transformation --- $S_m^+ = S_m^x + i S_m^y = (2S - b_{m}^{\dag} b_{m}^{})^{1/2} b_{m}^{}$, $S_m^- = S_m^x - i S_m^y = b_{m}^{\dag} (2S - b_{m}^{\dag} b_{m}^{})^{1/2}$, $S_{m}^{z} = S-b_{m}^{\dag} b_{m}^{}$, where $b_{m} (b_{m}^{\dag})$ is the bosonic magnon annihilation (creation) operator at site $m$. Within the low temperature approximation, we have $\left(2S-b^{\dag}b\right)^{1/2}\rightarrow\left(2S\right)^{1/2}$. Thus, retaining bilinear terms we get

\begin{eqnarray}
\begin{split}
&\mathcal{H} = h \sum_{m} b_{m}^{\dag} b_{m}^{}-\left[\left(\sum_{\langle mn \rangle_1} \left(J_{1} + i \nu_{mn} D_{1}\right) S b_{m}^{\dag} b_{n}^{}\right.\right.\\
&+\left.\left.\sum_{\langle \langle mn \rangle \rangle} i \nu_{mn} D_{2} b_{m}^{\dag} b_{n}^{}+ \sum_{\langle mn \rangle_2} J_{2} S b_{m}^{\dag} b_{n}^{} \right) + \text{H.c.}\right]  +E_{0}
\end{split}
\end{eqnarray} 
In the above, the ground state energy $E_0$ is not important. Thus, it will be regarded as zero. Next, we perform the Fourier transformation using the definition 
\begin{equation}
b_{\bm{k}}^{\dag}=\frac{1}{\sqrt{N}} \sum_{m} e^{i \bm{k} \cdot \bm{R}_m} b_{m}^{\dag}.
\end{equation}
Thus, in the reciprocal space the Hamiltonian is given by
\begin{equation}
\mathcal{H}=\sum_{\bm{k}} \psi_{\bm{k}}^{\dag} H(\bm{k}) \psi_{\bm{k}}^{},
\end{equation} where $\psi_{\bm{k}}^{\dag}=(b_{\bm{k}1}^{\dag},b_{\bm{k}2}^{\dag},b_{\bm{k}3}^{\dag},b_{\bm{k}4}^{\dag})$ and elements of the square-octagon Hamiltonian matrix are $
H_{ii}(\bm{k})= 2\nu_1 \cos \phi+\nu_2+h$,$
H_{12}(\bm{k})= -\nu_1 e^{i\bm{k}\cdot\bm{\delta}_1-i\phi}+i\nu_3\left(e^{i\bm{k}\cdot(\bm{\delta}_3+\bm{\delta}_4)}+e^{-i\bm{k}\cdot(\bm{\delta}_2+\bm{\delta}_3)}\right)$,$H_{13}(\bm{k})=-\nu_1 e^{-i\bm{k}\cdot\bm{\delta}_3+i\phi}-i\nu_3\left(e^{i\bm{k}\cdot(\bm{\delta}_1+\bm{\delta}_2)}+e^{i\bm{k}\cdot(\bm{\delta}_4-\bm{\delta}_1}\right)\notag$,$H_{24}(\bm{k})=-\nu_1 e^{-i\bm{k}\cdot\bm{\delta}_3-i\phi}+i\nu_3\left(e^{i\bm{k}\cdot(\bm{\delta}_1+\bm{\delta}_2)}+e^{i\bm{k}\cdot(\bm{\delta}_4-\bm{\delta}_1)}\right)$,$H_{34}(\bm{k})=-\nu_1 e^{i\bm{k}\cdot\bm{\delta}_1+i\phi}-i\nu_3\left(e^{-i\bm{k}\cdot(\bm{\delta}_2+\bm{\delta}_3)}+e^{i\bm{k}\cdot(\bm{\delta}_3+\bm{\delta}_4)}\right)\notag$,$H_{14}(\bm{k})=-i\nu_2 e^{i\bm{k}\cdot\bm{\delta}_4}$,$H_{23}(\bm{k})=-i\nu_2 e^{i\bm{k}\cdot\bm{\delta}_2}$.
The symbols are given by $\nu_1=\sqrt{J_1^2+D_1^2}S$, $\nu_2=J_2 S$, $\nu_3=D_2 S$, and $\phi=\arctan(J_1/D_1)$. The lattice vectors are given by $\bm{\delta}_1=(1-\sqrt{2}/2)(-1,1)a$, $\bm{\delta}_2=(\sqrt{2}-1)(-1,0)a$,  $\bm{\delta}_3=(1-\sqrt{2}/2)(-1,-1)a$, $\bm{\delta}_4=(\sqrt{2}-1)(0,-1)a$, where $a$ is the lattice constant. The energy bands are obtained by diagonalizing the bilinear spin-wave Hamiltonian. The results are shown in  Fig.~\ref{appdfig1}.
\begin{table}[h]
	\renewcommand\arraystretch{1.8}
	\centering
	\begin{tabular}{|c|c|c|}
		\hline Lattice &  Parameter & $T_c/J_1$ \\
		\hline \multirow{2}{*}{Square-octagon}  & $J_2, D_{1,2}/J_1=1, -0.6, 0$ &  0.254 \\
		\cline{2-3}
		& $J_2, D_{1,2}/J_1=1, 0, 0.3$ &  0.251 \\
		\hline Honeycomb & $D_{2}/J_1=0.3$ & 0.382 \\
		\hline \multirow{2}{*}{Kagome}  & $D_{1,2}/J_1=-0.6, 0$ &  0.527 \\
		\cline{2-3}
		& $D_{1,2}/J_1=0, 0.3$ &  0.534 \\ 			\hline
	\end{tabular} 
	\caption{\label{appdtab1} The disordering temperature $T_c$ in various lattices.}
\end{table}
Following exactly the same steps above, we can construct the Hamiltonian matrix of the kagome and the honeycomb lattices as follows
\begin{enumerate}
	\item {\bf Honeycomb lattice:} The model has nn ferromagnetic coupling $J_1$ and the nnn DM interaction $D_{2}$. The lattice vectors (also nnn vectors) are: $\bm{\alpha}_1=(1,0)a$, $\bm{\alpha}_2=(-1/2,\sqrt{3}/2)a$. Defining $\bm{\alpha}_3=-\bm{\alpha}_1-\bm{\alpha}_2$ and nn vectors $\bm{\beta}_1=(\bm{\alpha}_2-\bm{\alpha}_3)/3$, $\bm{\beta}_2=(\bm{\alpha}_3-\bm{\alpha}_1)/3$, $\bm{\beta}_3=(\bm{\alpha}_1-\bm{\alpha}_2)/3$, the elements of the Hamiltonian matrix are $H_{11}(\bm{k})= 3J_1 S+h+2D_{2}S\sum_{\delta=1}^{3}\sin(\bm{k}\cdot\bm{\alpha}_\delta)$,$H_{22}(\bm{k})= 3J_1 S+h-2D_{2}S\sum_{\delta=1}^{3}\sin(\bm{k}\cdot\bm{\alpha}_\delta)$,$H_{12}(\bm{k})= H_{21}^*(\bm{k}) = -2J_1 S\sum_{\delta=1}^{3}\exp(i \bm{k}\cdot\bm{\beta}_\delta)$.
	\item {\bf Kagome lattice:}~The model has nn ferromagnetic coupling $J_1$ and the nn and nnn DM interactions $D_{1}$, $D_{2}$, respectively. The lattice vectors are $\bm{a}_1=(1,0)a$, $\bm{a}_2=(-1/2,\sqrt{3}/2)a$. Defining $\bm{a}_3=-\bm{\alpha}_1-\bm{\alpha}_2$, nn vectors $\bm{\lambda}_i=\bm{a}_i$ and nnn vectors $\bm{\rho}_1=(\bm{\lambda}_3-\bm{\lambda}_2)$, $\bm{\rho}_2=(\bm{\lambda}_1-\bm{\lambda}_3)$, $\bm{\rho}_3=(\bm{\lambda}_2-\bm{\lambda}_1)$, the elements of the Hamiltonian matrix are: $H_{ii}(\bm{k})= 4 J_1 S+h$,$H_{12}(\bm{k})= -2(J_1-iD_{1})S  \cos(\bm{k}\cdot\bm{\lambda}_3)+2iD_{2}S\cos(\bm{k}\cdot\bm{\rho}_3)$,$H_{23}(\bm{k})= -2(J_1-iD_{1})S \cos(\bm{k}\cdot\bm{\lambda}_1)+2iD_{2}S\cos(\bm{k}\cdot\bm{\rho}_1)$,$H_{31}(\bm{k})= -2(J_1-iD_{1})S \cos(\bm{k}\cdot\bm{\lambda}_2)+2iD_{2}S\cos(\bm{k}\cdot\bm{\rho}_2)$.

\end{enumerate}
Besides, in realistic materials anisotropy terms can also exist(see Table.~\ref{appdtab3}), these terms are not included above though. The thermal disordering temperatures for the above lattices under the same parameter settings as our paper are provided in Table.~\ref{appdtab1}. The magnon plots in Figs. 4 and 5 of the main article are truncated at these temperature points.

\section{\label{Appd:B}Schwinger boson mean field theory}\label{appdB}
The spin-wave picture is effective in the low temperature regime. But, at high temperatures it is not. It is appropriate to consider the SBMFT (or bosonic spinon picture) to study the high temperature behavior of our model in this regime\cite{kim2016realization, lee2015thermal}. The spinon picture can be obtained by replacing the Holstein-Primakoff transformation by Schwinger boson transformation --- $\bm{S}_m= c_{ms}^{\dag}(\bm{\sigma})_{st}c_{mt}^{}/2$, where $\bm{\sigma}$ are Pauli matrices. The enlarged Hilbert space should be restricted by the constraint
\begin{equation}
\sum_{s}c_{ms}^{\dag}c_{ms}^{}=2S, s=\uparrow,\downarrow.
\end{equation}
The Hamiltonian has a quartic form after the transformation. Since, the full interacting problem is complicated to treat, we use a mean field method. First, we introduce a Lagrangian multiplier $\lambda_m$ for the constraint, then use the relations
\begin{eqnarray}
\bm{S}_{m}\cdot\bm{S}_{n}&=&\frac{1}{2}\sum_{st}\chi_{mns}^{\dag}\chi_{mnt}^{}-S(S+1),\\
(\bm{S}_{m}\times \bm{S}_{n})_z&=&-\frac{i}{2}\sum_{s} s\chi_{mns}^{\dag}\chi_{mn\bar{s}},
\end{eqnarray}
where $\chi_{mns}=c_{ms}^{\dag} c_{ns}^{}$. We then have

\begin{equation}
\begin{split}
\mathcal{H}&=-\frac{1}{2}\sum_{\langle mn \rangle_{1,2}}\sum_{st}J_{mn}\chi_{mns}^{\dag}\chi_{mnt}^{}\\
&-\frac{i}{2}\sum_{\langle mn \rangle_1 \langle \langle mn \rangle \rangle}\sum_{st} D_{mn}s\chi_{mns}^{\dag}\chi_{mn\bar{s}}^{} + \sum_{\langle mn \rangle_{1,2}}S(S+1)\\
&+\sum_{ms}\left(\lambda_m-\frac{sh}{2}\right)c_{ms}^{\dag}c_{ms}^{} - 2S\sum_{m}\lambda_m.
\end{split}
\end{equation}
Next, we define nine mean fields, namely --- the symmetric part of the $J_1$-type coupling $\zeta_{1s}$ and the corresponding antisymmetric part $\xi_{1s}$. Also for the $D_2$ type coupling, we define $\zeta_{3s}$ and $\xi_{3s}$, and the very last one is $\eta_2$ for the $J_2$-type coupling. Hence, we have
\begin{eqnarray}
\zeta_s&=&\braket{\chi_{mns}^{}+\chi_{mns}^{\dag}}/2,\\
\xi_s&=&\nu_{mn}\braket{\chi_{mns}^{}-\chi_{mns}^{\dag}}/2i,\\
\eta_2&=&\frac{1}{2}\sum_s \braket{\chi_{mns}^{}}=\frac{1}{2}\sum_s \braket{\chi_{mns}^{\dag}},
\end{eqnarray}
where we have assumed that $\lambda_m=\lambda$ for all sites. This reduces the Hamiltonian into a bilinear form by adopting the Hartree-Fock decoupling
\[\chi_{mns}^{\dag}\chi_{mnt}^{}\mapsto \braket{\chi_{mns}^{\dag}}\chi_{mnt}^{}+\chi_{mns}^{\dag}\braket{\chi_{mnt}^{}}-\braket{\chi_{mns}^{\dag}\rangle \langle\chi_{mnt}^{}}\]
Thus, the spinon Hamiltonian is now given by
\begin{equation}
\begin{split}
\mathcal{H}&= \left[\sum_{\langle mn \rangle_1 s}\left(-J_1\zeta_1  + J_1 \xi_1 i\nu_{mn}+\frac{D_1}{2} i\nu_{mn}s\zeta_{1\bar{s}}\right.\right.\\
&+\left.\frac{D_1}{2}s\xi_{1\bar{s}}\right)c_{ms}^{\dag}c_{ns}^{}- J_2\eta_2 \sum_{\langle mn \rangle_2 s}c_{ms}^{\dag}c_{ns}^{}\\
&+\left. \frac{D_2}{2}\sum_{\langle\langle mn \rangle\rangle s}\left(i\nu_{mn}s\zeta_{3\bar{s}}+s\xi_{3\bar{s}}\right)c_{ms}^{\dag}c_{ns}^{}+\text{H.c.}\right]\\
&+\sum_{ms}\left(\lambda-\frac{sh}{2}\right)c_{ms}^{\dag}c_{ms}^{}+F_0,
\end{split}
\end{equation}
where $\zeta_1=(\zeta_{1\uparrow}+\zeta_{1\downarrow})/2$, $\xi_1=(\xi_{1\uparrow}+\xi_{1\downarrow})/2$ and
\begin{equation}
\begin{split}
F_0&= 8J_1 N \left(\zeta_1^2+\xi_1^2\right)+(4J_1+2J_2) N S(S+1)\\
&- N\sum_s s\left(4D_1 \zeta_{1s}\xi_{1\bar{s}}+8D_2 \zeta_{3s}\xi_{3\bar{s}}\right)+4J_2 N \eta_2^2\\
&-8\lambda N S.
\end{split}
\end{equation}
\begin{figure*}[t]
	\centering
	\includegraphics[width=0.98\linewidth]{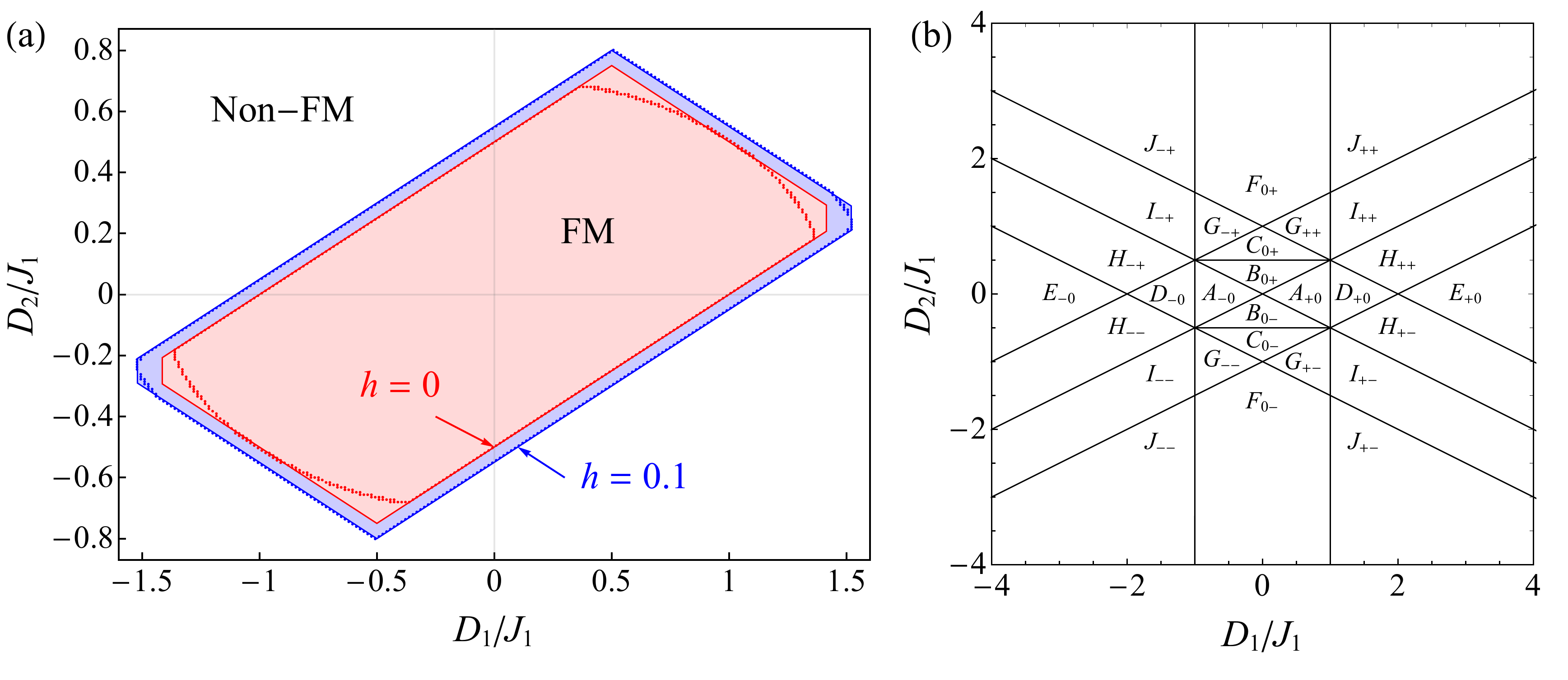}
	\caption{\label{appdfig2} Topological phase diagram in the $D_1 - D_2$ plane. (a)~Ferromagnetic phase region at $T=0$. The solid boundary denotes the ferromagnetic region restricted by high symmetry modes. The dotted one is the numerical calculation result over all modes. (b)~Topological phase diagram.The Chern number set for the various phases are given by $C_1$,~$C_2$,~$C_3$,~$C_4$: $A_{+0}$:$\{1,-1,-1,1\}$, $B_{0+}$:$\{-1,-1,1,1\}$, $C_{0+}$:$\{-1,3,-3,1\}$, $D_{+0}$:$\{-1,1,1,-1\}$, $E_{+0}$:$\{0,0,0,0\}$, $F_{0+}$:$\{0,2,-2,0\}$, $G_{++}$:$\{0,2,-3,1\}$, $H_{++}$:$\{0,0,1,-1\}$, $I_{++}$:$\{-2,0,3,-1\}$, $J_{++}$:$\{-2,1,3,-2\}$. $C_{\mu}(D_1,D_2)=-C_{\mu}(-D_1,-D_2)=-C_{5-\mu}(-D_1,D_2)$, $E_\mu<E_{\mu+1}$.}
\end{figure*}
If all mean fields are known, the spinon bands $E_{\mu\bm{k} s}$ can be obtained by diagonalizing this Hamiltonian. The free energy of the whole system is then given by
\begin{equation}
F=F_0+\frac{1}{k_B T}\sum_{\mu\bm{k} s}\ln [1-\exp\left(-E_{\mu\bm{k} s}/k_B T\right)],
\end{equation}
where $E_{\bm{k}\mu s}$ is the $n$th eigenenergy of spin $s$ with momentum $\bm{k}$.

\begin{table}[b]
	\renewcommand\arraystretch{1.8}
	\centering
	\begin{tabular}{|c|c|c|}
		\hline Mean Field & $\lambda_i(0)$/$J_1$ & $T_v$/$J_1$ \\
		\hline $\lambda$ & $2J_1 S + J_2 S + h/2=1.55$ & /  \\
		\hline $\zeta_{1\uparrow}$ & $2S=1$ & 0.81 \\ 
		\hline $\zeta_{1\downarrow}$ & 0 & 0.81 \\
		\hline $\zeta_{3\uparrow}$ & $2S=1$ & 0.71 \\ 
		\hline $\zeta_{3\downarrow}$ & 0 & 0.71 \\
		\hline $\eta_{2}$ & $S=0.5$ & 0.71 \\ 
		\hline $\xi_{1s}$, $\xi_{3s}$ & 0 &/\\ 	\hline
	\end{tabular} 
	\caption{\label{appdtab2} The zero-temperature value $\lambda_i(0)$ and the vanishing temperature $T_v$ of mean fields. Parameters are: $J_1=J_2=1$, $h=0.1$, $D_1=0.6$, $D_2=0.1$.}
\end{table}

The mean fields were determined by minimizing the free energy. This gives us
\begin{equation}
\frac{\partial F}{\partial \lambda_i}=\frac{\partial F_0}{\partial \lambda_i}+\sum_{\mu\bm{k} s}n_B(E_{\mu\bm{k} s})\frac{\partial E_{\mu\bm{k} s}}{\partial \lambda_i}=0,
\end{equation}
where $n_B(\varepsilon)=(\exp(\varepsilon/k_B T)-1)^{-1}$ is the Bose distribution and $\lambda_i\in(\lambda,\eta_2,\zeta_{1s},\xi_{1s},\zeta_{3s},\xi_{3s})$. Those equations are self consistency equations of the mean fields. At zero temperature, the up spinon condensates and the down spinon will be gapped by the magnetic field, thus the self-consistent equations will be given by
\begin{equation}
\frac{\partial F_0}{\partial \lambda_i}+8NS\frac{\partial E_{1\bm{0} \uparrow}}{\partial \lambda_i}=0.
\end{equation}
The zero-mode energy of the first band of the up spinon is
\begin{equation}
E_{1\bm{0}\uparrow}=\lambda-2J_1\zeta_1-J_2\eta_2-\frac{h}{2}+D_1 \xi_{1\downarrow}+2D_2 \xi_{3\downarrow}.
\end{equation}
Solving the above equations we can get the zero-temperature value of the mean fields as
\begin{equation}
2\eta_2=\zeta_{1\uparrow}=\zeta_{3\uparrow}=2S,~\xi_{1s}=\xi_{3s}=\zeta_{1\downarrow}=\zeta_{3\downarrow}=0.
\end{equation}
The Lagrangian multiplier is determined by the condensation condition
\begin{equation}
E_{1\bm{0}\uparrow} = 0 \Rightarrow \lambda = (2J_1+J_2)S+h/2.
\end{equation}

Comparing the zero-temperature Hamiltonian of the spinons to the magnon Hamiltonian, we have $H_{\downarrow}=H_{\text{mag}}$, and $H_{\uparrow}=H_{\text{mag}}(D_{1,2}=0)$, which explains the consistency of low temperature thermal Hall conductance results in our paper. In Table.~\ref{appdtab2} we list the temperature at which the mean field value vanishes within the same parameter setting of our paper.

\section{\label{Appd:C}Magnetic and topological phase diagrams}\label{appdC}
The ferromagnetic phase is an eigenstate
\begin{eqnarray}
\ket{{\Psi_0}}&=&\bigotimes_{m}\ket{{\uparrow}_m},\notag\\
\mathcal{H}\ket{{\Psi_0}}&=&-\left(\frac{1}{2}(2J_1+J_2)N+2 h N\right)\ket{{\Psi_0}}.
\end{eqnarray}
Hence, if the bottom of the lowest band is the zero-mode, then definitely the ground state (at zero temperature) is ferromagnetic. In fact, the analytical expressions for the energies at high symmetry points are given by
\begin{eqnarray}
E(0,0)&=& h+2S\left\{0,2J_1,J_1+J_2\pm (D_1+2D_2)\right\},\notag\\
E(\pi,0)&=& h+S\left\{2J_1+J_2\pm\sqrt{4D_1^2+J_2^2},\right.\notag\\
&& \left.2J_1+J_2\pm\sqrt{4J_1^2+J_2^2}\right\},\\
E(\pi,\pi)&=& h+2S\left\{J_2,2J_1+J_2,J_1\pm (D_1-2D_2)\right\}.\notag	
\end{eqnarray}
\begin{figure*}[t]	
	\centering
	\includegraphics[width=\linewidth]{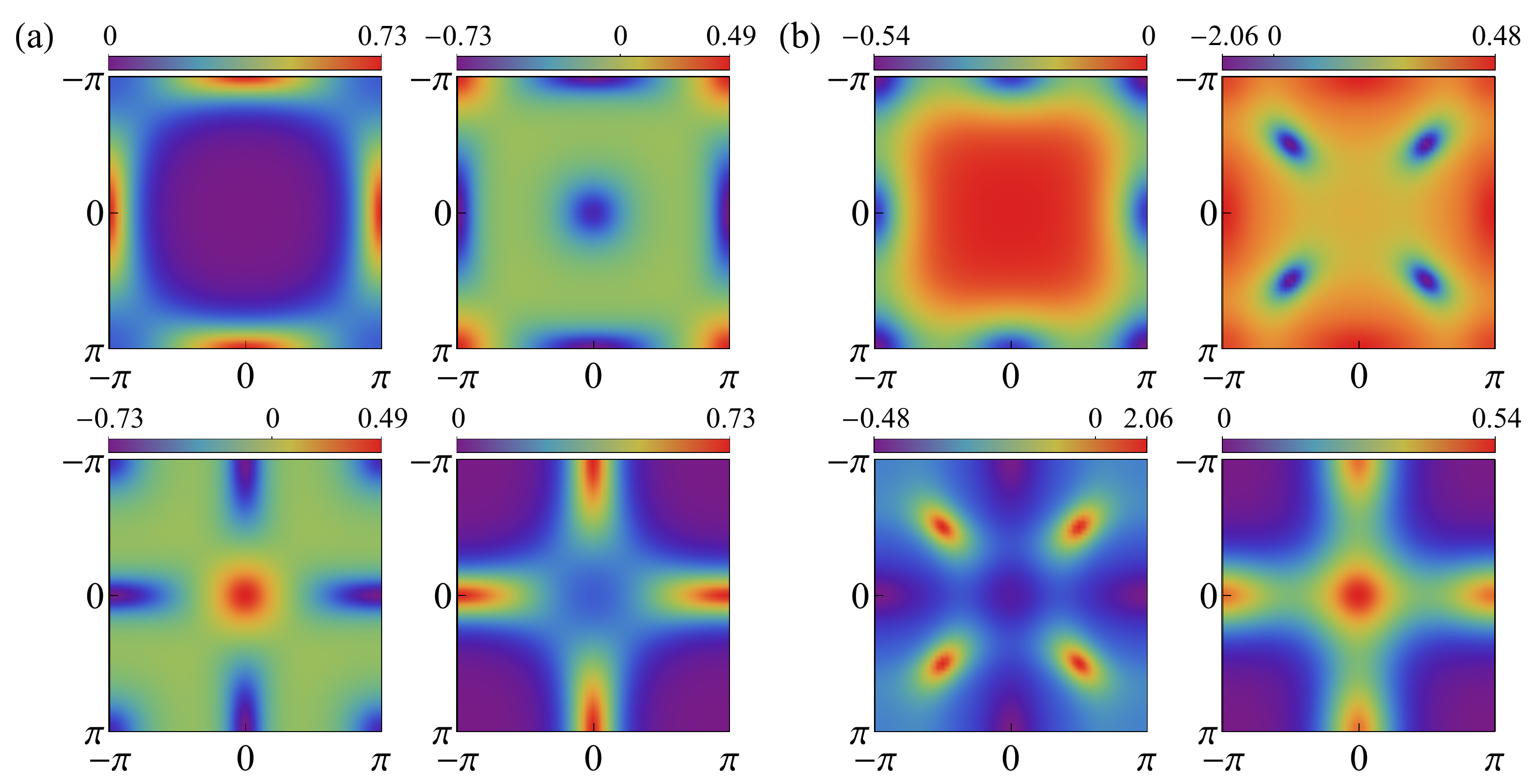}
	\caption{\label{appdfig3} Berry curvature of magnon bands with nonzero DM interactions. The upper left, upper right, lower left, lower right figure correspond to the first, second, third, and fourth band (from lower to higher) respectively. Parameters: (a)~$S=1/2$, $J_1=J_2=1.0$, $D_1=0.6J_1$, $D_2=0$; (b)~$S=1/2$, $J_1=J_2=1.0$, $D_1=0$, $D_2=0.3J_1$.}
\end{figure*}
The band gap introduced by DM interactions is clearly from the above equations, that is $\pm 2(D_1+2D_2)S$ at $(\pi,0)$ and $\pm 2|D_1-2D_2|S$ at $(\pi,\pi)$, see Fig.~\ref{appdfig1}. By comparing the energies at those points, we can get a rough ferromagnetic phase region in $D_1-D_2$ parametric space while $J_1=J_2=1, S=1/2$.
\begin{eqnarray}
|D_1-2D_2|&\leqslant& 1+h\notag\\
|D_1+2D_2|&\leqslant& 2+h\\
|D_1|&\leqslant& \sqrt{(h+1)(h+2)}.\notag
\end{eqnarray}
In Fig. 1b, we reported only a small portion (upper right hand corner) of the magnetic phase diagram of the square-octagon lattice. In Fig.~\ref{appdfig2}, we present our extensive mapping of the ferromagnetic phase over all energy modes by numerical calculation. We demarcate the non-ferromagnetic phases from the ferromagnetic zones, only. We have not tried to analyze the non-ferromagnetic phases, since it is outside the scope of our present work.

Topological phases are distinguished by Chern numbers of energy bands. As an analog of the magnetic field in quantum Hall states, DM interactions are important to introduce nonzero Berry curvatures. The equations of them are given by Eqs.~(\ref{BerryCurvature}) and (\ref{ChernNumber}). We show Berry curvature of magnon bands in Fig.~\ref{appdfig3}. The Berry curvature can be positive or negative. This offers an explanation of the sign change behavior of thermal Hall conductance. As topological invariants, Chern numbers are integers, and will change only when the topological structure of energy bands change, that is, when the two bands degenerate \cite{hatsugai1993edge, thouless1982quantized}. We can find all the degenerate conditions by solving the multiple root discriminant of the characteristic polynomial of the Hamiltonian matrix, which has  a quartic form. However, it is possible that sometimes the two bands are degenerate with conserved Chern numbers, resulting in a trivial degeneration. So we numerically checked the condition near each boundary between different phases carefully and finally got the topological phase diagram (Fig.~\ref{appdfig2}).

\section{\label{Appd:D}Edge states in a strip sample}\label{appdD}

\begin{figure}[b]
	\centering
	\includegraphics[width=0.98\linewidth]{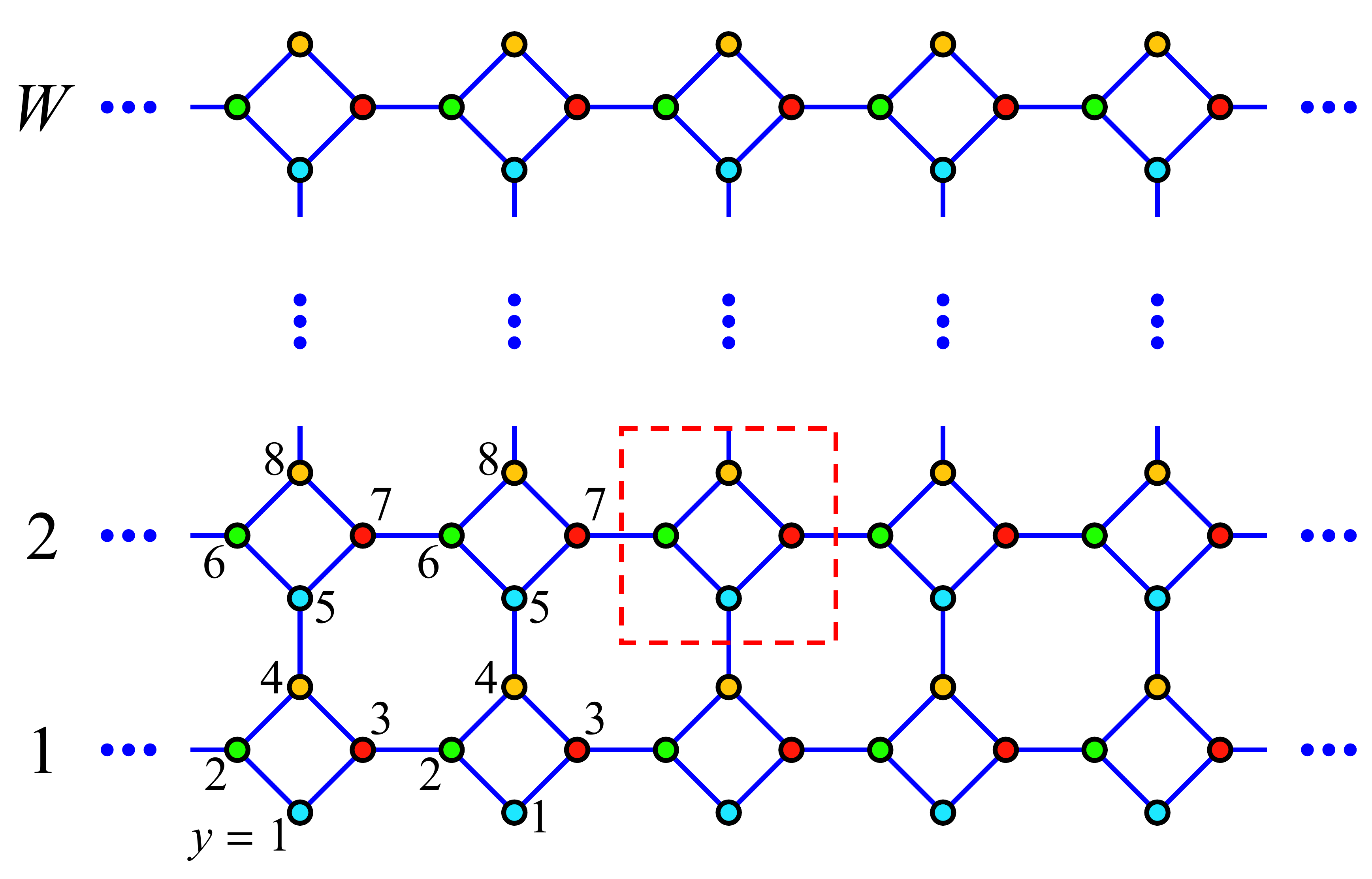}	
	\caption{\label{appdfig4} Strip Sample. The sample has $W$ periodic one-dimensional chains, the numbers nearing sites are $y$ indices.}
\end{figure}

To compute the edge states, we considered a strip sample, expanded to infinity along the $x$ direction but finite along $y$ direction, and with two edges (the top edge and the bottom edge) perpendicular to $y$ direction as shown in Fig.~\ref{appdfig4}. In this case, $k_y$ is not a good quantum number any more. We need to designate those one-dimensional chains whose locations share the same $y$ coordinate and belong to the same sublattice as $y$. Then we can rewrite the Hamiltonian in the $(k_x, y)$ space.

For simplicity, we replace $k_x$ by $k$, and let $y$ run from $i_1$ to $4(W-1)+i_2$ $(i_1, i_2 \in \{1, 2, 3, 4\})$, where $W$ denotes the number of periodic one-dimensional chains along the $y$-direction.
\begin{equation}
b_{ky}^{\dag}=\frac{1}{\sqrt{N_x}} \sum_{m} e^{i k \bm{R}_m \cdot \bm{e}_x} b_{my}^{\dag}.
\end{equation}
The new Hamiltonian is
\begin{equation}
\mathcal{H}=\sum_{k} \varphi_{k}^{\dag} H(k) \varphi_{k}^{},
\end{equation}
where $\varphi_k^{\dag} = (b_{k, i_1}^{\dag}, b_{k, i_1+1}^{\dag}, \dots, b_{k, 4(W-1)+i_2}^{\dag})$. While $i_1=1$, $i_2=4$, under open boundary condition $b_{k,0}^{\dag}\ket{\text{Ground State}}=b_{k,4W+1}^{\dag}\ket{\text{Ground State}}=0$, the Hamiltonian should be
\begin{equation}
H(k)=\begin{pmatrix}
G(k)  & F^{\dag}(k)& 0          & \cdots& 0\\
F(k)  & G(k)       & F^{\dag}(k)& \ddots& \vdots\\
0     & F(k)       & \ddots     & \ddots& 0\\
\vspace{6pt}
\vdots& \ddots     & \ddots     & \ddots& F^{\dag}(k)\\
0     & \cdots     & 0          & F(k)  & G(k) 
\end{pmatrix}_{4W \times 4W}
\end{equation}
where $G$ and $F$ are $4\times 4$ matrices with $G_{ii}=2\nu_1 \cos \phi+\nu_2+h$, $G_{ij}(k)=G_{ji}^{*}(k)$, $G_{14}=0$, $G_{12}=G_{13}^{*}=-\nu_1 e^{-i\theta_1-i\phi}+i\nu_3 e^{i\theta_2}$, $G_{24}=G_{34}^{*}=-\nu_1 e^{i\theta_1-i\phi}+i\nu_3 e^{-i\theta_2}$, $G_{23}=-\nu_2 e^{-2i\theta_1}$,  $F_{24}=-F_{13}=F_{34}^{*}=-F_{12}^{*}=i\nu_3 e^{i\theta_1}$,  $F_{14}=-\nu_2$, $F_{ij}=0\ (\text{otherwise})$,  $\theta_1(k)=(1-\sqrt{2}/2)ka$, $\theta_2(k)=(\sqrt{2}/2)ka$. If we want to set different $i_1$, $i_2$ to adjust edge condition, we can just cut the Hamiltonian matrix into a smaller matrix, where the first (last) column (row) is the $i_1$th ($4(W-1)+i_2$ th) column (row) of the previous matrix. We choose $W=120$ which ensures that the computed edge state features have converged and are insensitive to further strip width size increase.

\section{\label{Appd:E}Thermal Hall conductance and angular momentum}\label{appdE}

We can calculate the transport properties of magnons based on linear response theory, according to the well-known papers of Matsumoto and Murakami\cite{matsumoto2011rotational, matsumoto2011theoretical}. The thermal Hall conductance of magnons is given by Eq.~(\ref{ThermalHallConductance}). For SBMFT, we need to perform an additional summation over the spin index.

The total angular momentum per unit cell has two components, the edge current and the self-rotation
\begin{equation}
L_{\text{tot}}=m^* l_{\text{tot}}, l_{\text{tot}}=l_{\text{edge}}+l_{\text{self}},
\end{equation}	
where $m^*$ is the effective mass of the first band at $\Gamma(0,0)$, within low temperature approximation. And these parts have the form
\begin{equation}
\begin{split}
l_{\text{tot}}&=\frac{4}{\hbar} \frac{1}{N} \text{Im} \sum_{\mu\bm{k}} \Bra{\frac{\partial \psi_\mu}{\partial k_x}} k_B T c_1(n_B(\varepsilon_{\mu\bm{k}}))\\
&-\frac{n_B(\varepsilon_{\mu\bm{k}})}{2}(H+\varepsilon_{\mu\bm{k}})\Ket{\frac{\partial \psi_\mu}{\partial k_y}}
\end{split}
\end{equation}
and
\begin{equation}
l_{\text{self}}=\frac{4}{\hbar} \frac{1}{N}\text{Im} \sum_{\mu\bm{k}} \Bra{\frac{\partial \psi_\mu}{\partial k_x}} \frac{n_B(\varepsilon_{\mu\bm{k}})}{2}(\varepsilon_{\mu\bm{k}}-H)\Ket{\frac{\partial \psi_\mu}{\partial k_y}},
\end{equation}
where $c_1(x)=(1+x)\ln{(1+x)}-x \ln{x}$ is another weight function.

\begin{figure}[h]
	\centering
	\includegraphics[width=0.98\linewidth]{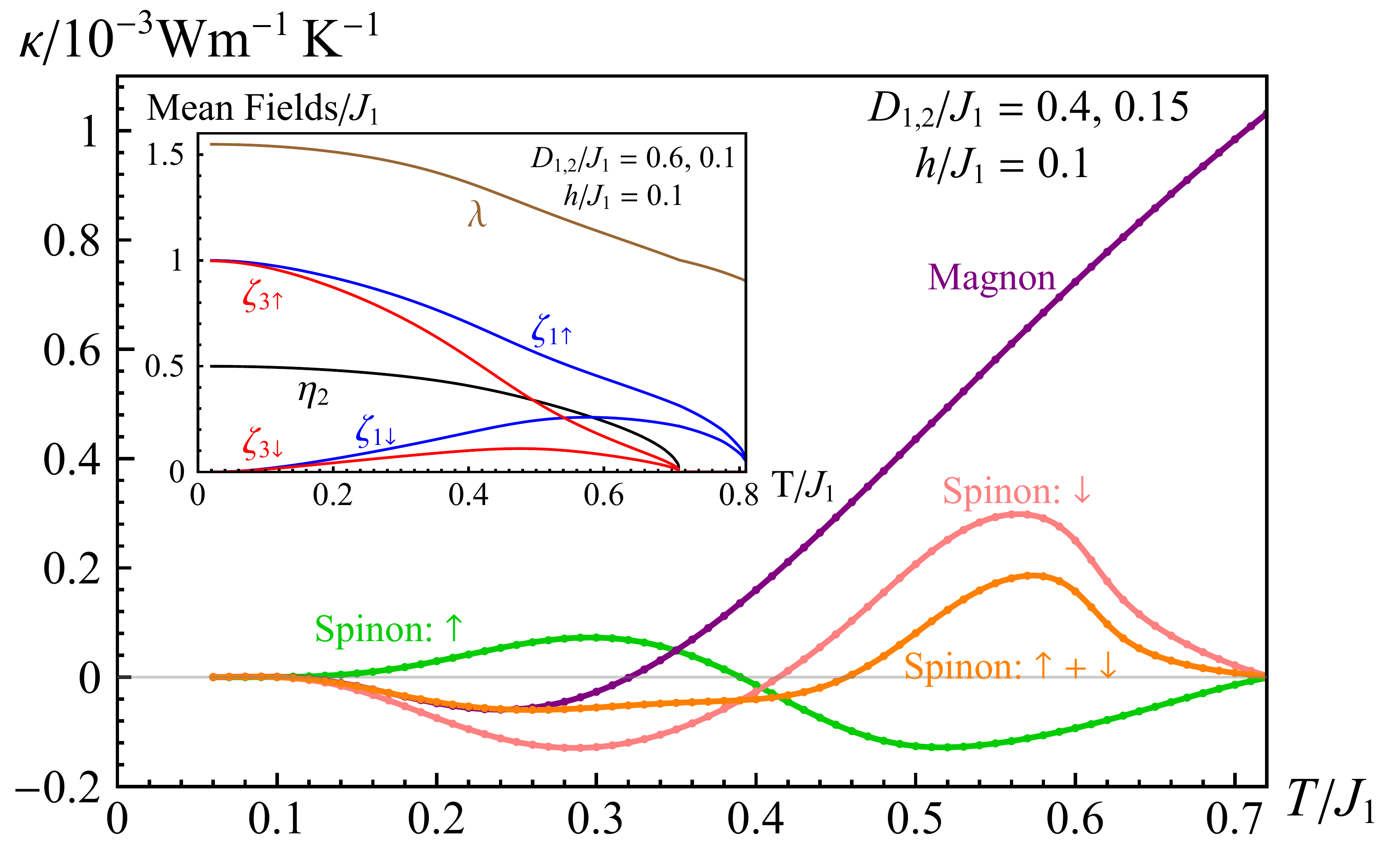}
	\caption{\label{appdfig5} Thermal Hall $\kappa_{xy}$ response of spinon and magnon fields. Inset picture shows converged values of spinon mean field parameters. The magnon curve is terminated at the critical thermal disordering temperature.}
\end{figure}
\begin{table*}[t]
	\renewcommand\arraystretch{1.8}
	\centering
	\begin{tabular}{|c|c|c|}
		\hline Material & $\text{CrI}_3$ & $\text{Cu(1,3-bdc)}$ \\
		\hline Hamiltonian $\mathcal{H} = \sum_{m<n} [ J_{mn} \bm{S}_{m}\cdot\bm{S}_{n}+D_{mn}\cdot (\bm{S}_{m}\times \bm{S}_{n})_z]$ & $- \sum_{m} A_z (S_m^z)^2 $ & $-g \mu_{B} H \sum_{m} S_m^z$  \\
		\hline Lattice & Honeycomb & Kagome \\
		\hline Lattice constant $a$~(nm)& $\approx 0.686$ & $\approx 0.910$\\
		\hline Spin & $S=3/2$ & $S=1/2$ \\ 
		\hline Z anisotropy or g factor and magnetic field& $A_z=0.22~ \text{meV}$ & $g=2.2$,~$\mu_B H = 7~\text{T}$ \\
		\hline \multirow{3}{*}{Heisenberg Interaction $J$ ($\text{meV}$)} & 
		$J_{1}=2.01$
		& \multirow{3}{*}{$J_{1}=0.6$} \\
		& $J_{2}=0.16$&\\
		& $J_{3}=-0.08$&\\
		\hline DM Interaction $D$ ($\text{meV}$) & $D_{2}=0.31$ & $D_1= 0.09$\\
		\hline Curie Temperature $T_c$~(K) & $\approx 61$ & $\approx 1.8$ \\ \hline
	\end{tabular}
	\caption{\label{appdtab3} Model and parameters of two kinds of two-dimensional magnon topological insulators.}
\end{table*}
Here we give out the calculation result of thermal Hall conductance of both magnons and spinons. The mean fields are already defined in the previous sections. Among the ten mean fields, four anti-symmetric parts are not important. The result is displayed in the inset of Fig.~\ref{appdfig5}. From the figure, we can see that all mean fields vanish above $T_c \approx 0.81 J_1$, the boundary of the ferromagnetic-paramagnetic phase transition. In the zero temperature limit and with an external nonzero field $h>0$, the mean fields tend to the following limits  $\lambda\rightarrow 2J_1 S+J_2 S+h/2$, $\zeta_{1\uparrow}=\zeta_{3\uparrow}\rightarrow2S$, $\eta_2\rightarrow S$ and $\zeta_{1\downarrow}=\zeta_{3\downarrow}=\xi_{1\uparrow}=\xi_{1\downarrow}=\xi_{3\uparrow}=\xi_{3\downarrow}=0$ in the ferromagnetic phase region. Thus, the Hamiltonian of the spin down is the same as the magnon $H_{\downarrow}=H_{\text{mag}}$~\cite{kim2016realization}. However, the Hamiltonian of the spin up corresponds to a topologically trivial $H_{\uparrow}=H_{\text{mag}}(D_1=D_2=0)$. The focus of the present work is on the fully magnetized regime of the sample. At low temperature, where the spinons condense into magnons, the results from the magnon and the spinon calculations agree. This is a check on the validity of our calculation from two separate formalism. Since the spinon picture coincide with magnons at low temperature, and Figs.~2(c) and 2(d) shows another similarity between angular momentum of magnons and thermal Hall conductance, we do anticipate spinons to exhibit EdH effect at higher temperatures.

\section{\label{Appd:F}NEGF method}\label{appdF}

In this section we introduce the nonequilibrium Green's function (NEGF) method for calculating transport properties of magnons. The square-octagon lattice sample was divided into three zones, the left part, the central part, and the right part. The temperature of the left and the right part was kept at $T_L$ and $T_R$. The magnon transport of the central part is what we were concerned with. The full Hamiltonian of this system is given by Eq.(\ref{NEGFHamiltonian}) \cite{zhang2013topological}.

In the interaction picture, $H_{LL,CC,RR}$ was treated as a free Hamiltonian for each part, while $H_{LC}=H_{CL}^{\dag}$, $H_{CR}=H_{RC}^{\dag}$ were interactions. In the non-equilibrium case, we need to use the contour-ordered Green's function, which is defined on a contour near the interval $(-\infty,t)$ of the real axis of the time complex plane.

In the non-equilibrium case, we need to use the contour-ordered Green's function, which is defined on a contour near the interval $(-\infty,t)$ of the real axis of the time complex plane. It gives the same result as the real-time Green's function in non-equilibrium case but can be used also in non-equilibrium case \cite{haug2009quantum}.

The contour-ordered Green's function is defined as
\begin{equation}
G^{\sigma\sigma^{\prime}}(t,t^{\prime})=-i\langle T_C b(\tau_{\sigma}) b^{\dag}(\tau_{\sigma^{\prime}}^{\prime})\rangle,\ \tau_{\pm}=t\pm i 0^+.
\end{equation}
For nonequilibrium steady state, we have $\tilde{G}(\tau,\tau^{\prime})=\tilde{G}(\tau-\tau^{\prime})$.
There are some relations between contour-ordered Green's functions and real-time Green's functions
\begin{equation}
G^{++}=G^t,\ G^{--}=G^{\bar{t}},\ G^{+-}=G^>,\ G^{-+}=G^<.
\end{equation}
The retarded and advanced Green's function can also be expressed as
\begin{eqnarray}
G^r&=& G^t-G^<=G^>-G^{\bar{t}},\notag\\
G^a&=& G^t-G^>=G^<-G^{\bar{t}}.
\end{eqnarray}
Without interaction, we have
\begin{equation}
g_{\alpha}^r (\varepsilon)=\left[(\varepsilon +i \eta)-H_{\alpha \alpha}\right]^{-1}\text{, }\ g_{\alpha}^a(\varepsilon)=\left[g_{\alpha}^r(\varepsilon)\right]^{\dag}.
\end{equation}

For the equilibrium system, there's another relation \cite{zhang2013topological}
\begin{equation}
g_{\alpha}^< (\varepsilon)=n_{\alpha}(\varepsilon)\left(g_{\alpha}^a (\varepsilon)-g_{\alpha}^r (\varepsilon)\right),
\end{equation}
where $n_{\alpha}(\varepsilon)=\langle b^{\dag}b\rangle=(e^{\varepsilon/T_{\alpha}}-1)^{-1}$ is the Bose function, and we take $\hbar=k_B=1$. With interactions present, we have the Dyson equation
\begin{equation}
((\varepsilon+i\eta)-H)\begin{pmatrix}
G_{LL}^r(\varepsilon)& G_{LC}^r(\varepsilon)& G_{LR}^r(\varepsilon)\\
G_{CL}^r(\varepsilon)& G_{CC}^r(\varepsilon)& G_{CR}^r(\varepsilon)\\
G_{RL}^r(\varepsilon)& G_{RC}^r(\varepsilon)& G_{RR}^r(\varepsilon)
\end{pmatrix}=I.
\end{equation}
Solving the above equation we get \cite{wang2008quantum}
\begin{equation}
G_{CC}^r (\varepsilon)=[G_{CC}^a(\varepsilon)]^{\dag}=[\varepsilon+i\eta-H_{CC}-\Sigma^r (\varepsilon)]^{-1},
\end{equation}
where the self energy is
\begin{equation}
\Sigma^{r,<} (\varepsilon)=H_{CL}g_L^{r,<} H_{LC}+H_{CR}g_R^{r,<} H_{RC},
\end{equation}
the lesser self energy has the same form as the retarded self energy because of the Langreth theorem \cite{haug2009quantum}. We can get $G^<$ through Keldysh equation \cite{haug2009quantum}
\begin{equation}
G_{CC}^<(\varepsilon)=G_{CC}^r(\varepsilon)\Sigma^<(\varepsilon) G_{CC}^a(\varepsilon).
\end{equation}
With the above Green's functions we can calculate the local density and current of this system as Eqs.~(\ref{NEGFDensity}) and (\ref{NEGFCurrent}).

\section{\label{Appd:G}Parameters of realistic materials}\label{appdG}

We calculated the size of Einstein-de Haas effect in two kinds of two-dimensional materials --- $\text{CrI}_3$ and $\text{Cu(1,3-bdc)}$. The Hamiltonian of them have some differences, but can still be treated within magnon picture. The parameters of these models are listed in Table.~\ref{appdtab3}.

\vspace{4mm}
\begin{acknowledgments}
T.~D. acknowledges funding support from Sun Yat-Sen University Grants No.~OEMT-2017-KF-06 and No. OEMT-2019-KF-04. J.~ L. and D.~X.~Y. are supported by Grants No. NKRDPC-2017YFA0206203, No. NKRDPC-2018YFA0306001, No. NSFC-11974432, No. GBABRF-2019A1515011337, and the Leading Talent Program of Guangdong Special Projects.
\end{acknowledgments}

\bibliography{PRR}

\end{document}